\documentclass{aa}

\usepackage{graphicx,natbib,txfonts,color}
\usepackage{bm}
\usepackage{multicol, multirow}
\bibpunct{(}{)}{;}{a}{}{,}

\newcommand{\feh}{\textrm{[Fe/H]}}
\newcommand{\teff}{T_{\mbox{\textrm{\scriptsize eff}}}}
\newcommand{\gk}{{G-K}_{\mbox{\textrm{\scriptsize s}}}}
\newcommand{\ks}{{K}_{\mbox{\textrm{\scriptsize s}}}}
\newcommand{\hp}{{H}_{\mbox{\textrm{\scriptsize p}}}}
\newcommand{\tnorm}{\hat{T}}
\def\arcsec{\,$''$}

\begin{document}


\title{Empirical photometric calibration of the Gaia Red Clump: \\ colours, effective temperature and absolute magnitude \thanks{Table \ref{tab:hrtgas} is only available in electronic form at the CDS}}

\author{L. Ruiz-Dern, C. Babusiaux, F. Arenou, C. Turon, R. Lallement}

\offprints{laura.ruiz-dern@obspm.fr}

\institute{GEPI, Observatoire de Paris, PSL Research University, CNRS UMR 8111 - 5 Place Jules Janssen, 92190 Meudon, France}

\date{Received 17 July 2017; Accepted 16 October 2017}

\authorrunning{Ruiz-Dern et al.}
\titlerunning{Gaia Red Clump photometric calibration}

\abstract
{Gaia Data Release 1 allows to recalibrate standard candles such as the Red Clump stars. To use those, they first need to be accurately characterised. In particular, colours are needed to derive the interstellar extinction. As no filter is available for the first Gaia data release and to avoid the atmosphere model mismatch, an empirical calibration is unavoidable.
}
{The purpose of this work is to provide the first complete and robust photometric empirical calibration of the Gaia Red Clump stars of the solar neighbourhood, through colour-colour, effective temperature-colour and absolute magnitude-colour relations, from the Gaia, Johnson, 2MASS, Hipparcos, Tycho-2, APASS-SLOAN and WISE photometric systems, and the APOGEE DR13 spectroscopic temperatures.
}
{We used a 3D extinction map to select low reddening red giants. To calibrate the colour-colour and the effective temperature-colour relations, we developed a MCMC method which accounts for all variable uncertainties and selects the best model for each photometric relation. We estimate the Red Clump absolute magnitude through the mode of a kernel-based distribution function.
}
{We provide 20 colour vs $\gk$ relations and the first $\teff$ vs $\gk$ calibration. We obtained the Red Clump absolute magnitudes for 15 photometric bands with, in particular, $M_{\ks} = (-1.606 \pm 0.009)$ and $M_G = (0.495 \pm 0.009) + (1.121 \pm 0.128) \; (\gk-2.1)$. We present an unreddened Gaia-TGAS HR diagram and use the calibrations to compare its Red Clump and its Red Giant Branch Bump with the Padova isochrones.
}
{}


\maketitle

\graphicspath{{Figures/}{/home/laura/Documents/00_PhD/Work/Papers/07_CalibrationRC/Figures/}}

\section{Introduction}\label{sect:intro}

Measuring distances with high accuracy is as difficult as fundamental in astronomy. The most direct method for estimating astronomical distances is the trigonometric parallax. However, relative precisions of parallaxes decrease with distance. In order to go further we need to use standard candles such as Red Clump (hereafter RC) stars. 

RC stars are low mass core \textit{He}-burning (CHeB) stars and cooler than the instability strip. They appear as an overdensity in the Colour-Magnitude Diagram (CMD) of populations with ages older than $\sim 0.5 - 1$Gyr, covering the range of spectral types G8III - K2III with 4500K $\lesssim \teff \lesssim$ 5300K. Indeed, the RC represents the young and metal-rich counterpart of the Horizontal Branch \citep[see ][ for a review]{girardi_rcstars_2016}.

The Red Clump is used as a standard candle for estimating astronomical distances due to its relatively small dependency of the luminosity on the stellar composition, colour and age in the solar neighbourhood \citep{paczynski_galactocentric_1998, stanek_distance_1998, udalski_optical_2000, alves_k-band_2000, groenewegen_red_2008, valentini_spectroscopic_2010}. As stated by \cite{paczynski_galactocentric_1998}, any method to obtain distances to large-scale structures suffers mainly from four problems: the accuracy of the absolute magnitude determination of the stars used, the interstellar extinction, the distribution of the inner properties of these stars (mass, age, chemical composition), and how large the sample is. Whether the use of the RC may be considered particularly different than other standard candles such as RR Lyrae or Cepheids, is precisely due to their large number. The larger the sample used, the lower the statistical error in distance calculations. To efficiently use the Red Clump as a standard candle, a good characterisation of the calibrating samples, here the solar neighbourhood, is needed, to which stellar population corrections can then be applied \citep[see e.g.][]{girardi_fine_1998}.

The First Gaia Data Release (GDR1) was delivered to the scientific community in September 2016 \citep{gaia_dr1summary_2016, gaia_dr1_2016}. Although we will have new and more accurate astrometric and photometric measurements for thousands of RC stars in future releases, this first catalogue includes the Tycho-Gaia Astrometric Solution (TGAS) subsample \citep{lindegren_gdr1_2016} with already a significant set of accurate solar neighbourhood RC parallaxes (with systematic error at the level of 0.3 mas, i.e. 3 times better than in the Hipparcos catalogue, for 20 times more stars).  

\begin{figure*}
\centering
\begin{tabular}{cc}
\includegraphics[width=0.45\textwidth]{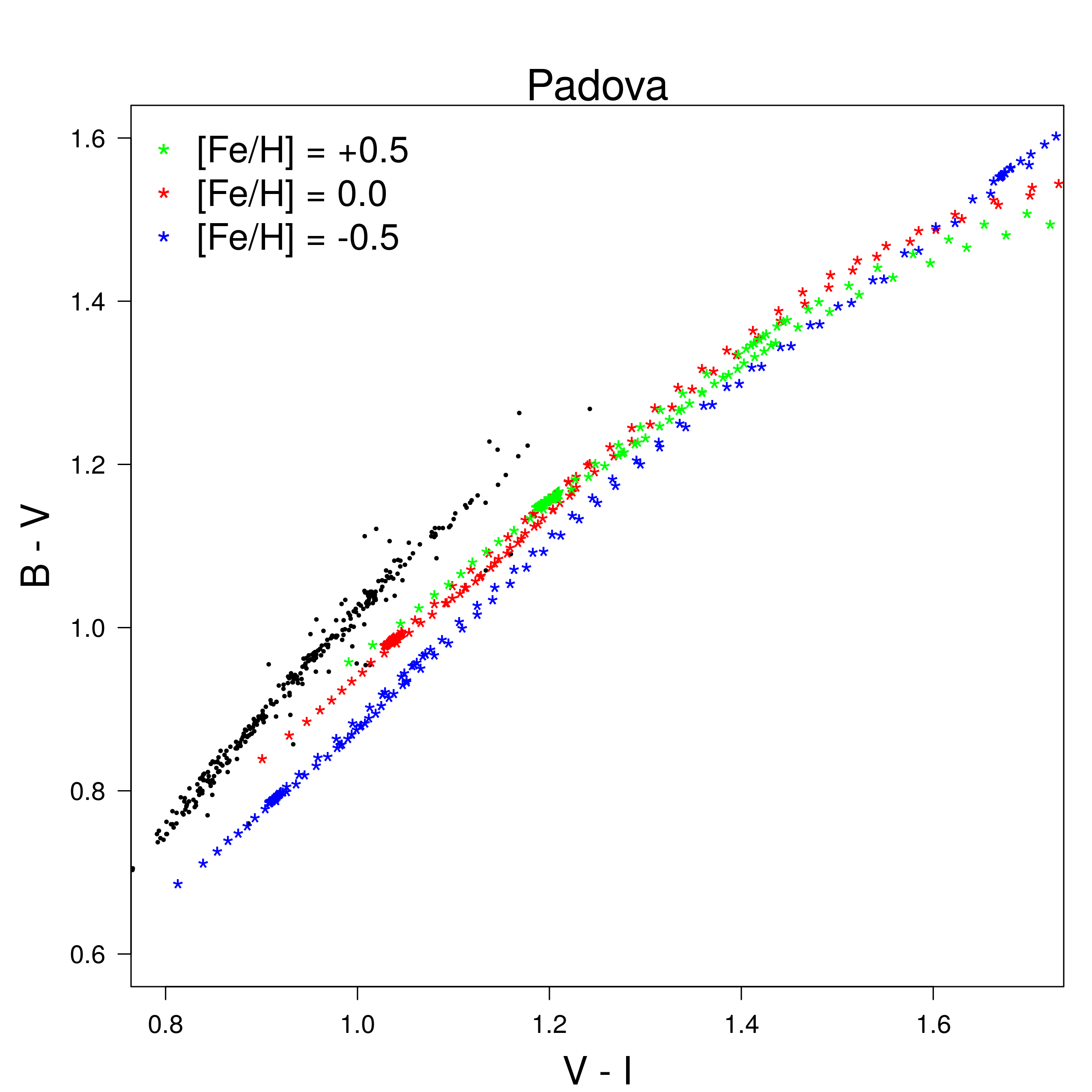} &
\includegraphics[width=0.45\textwidth]{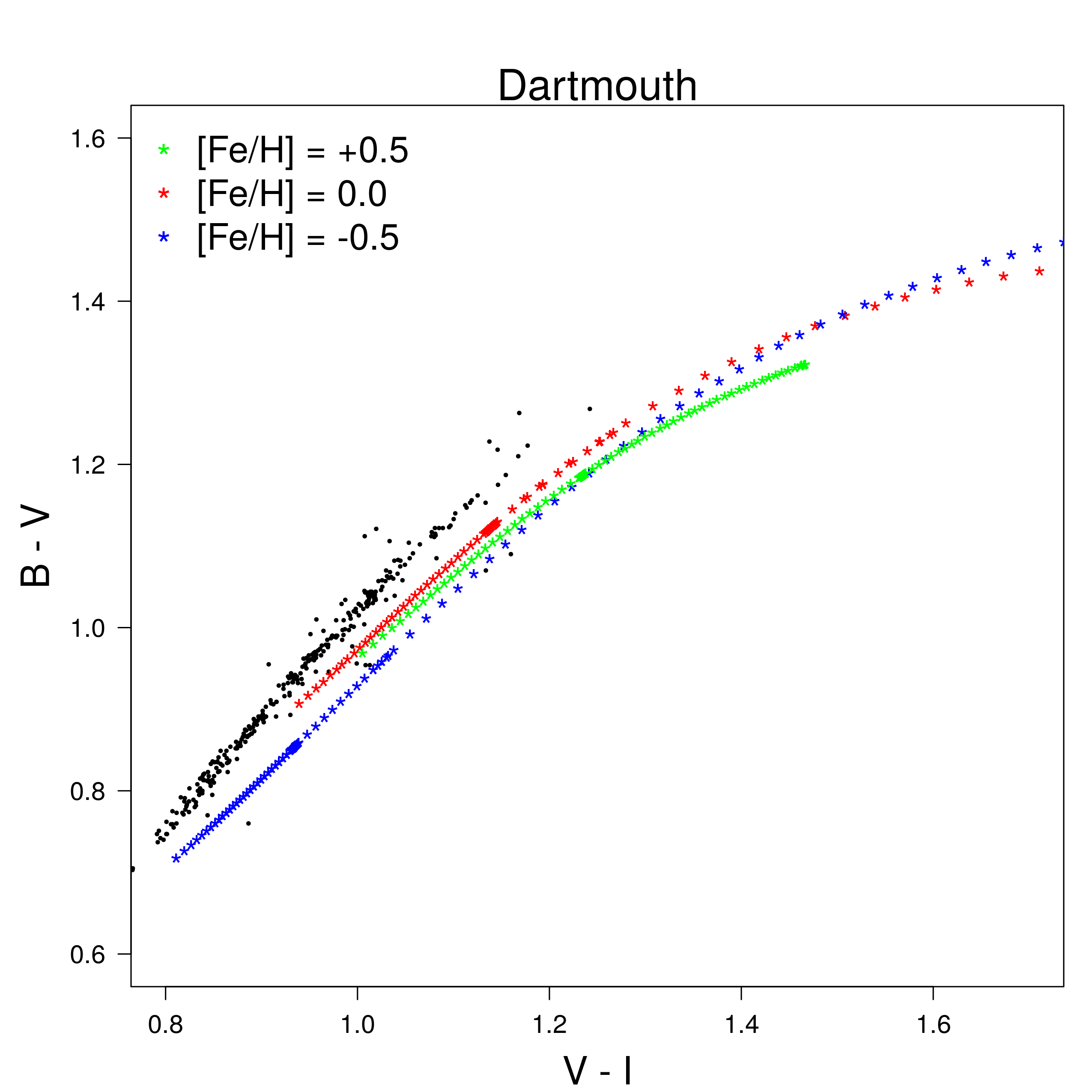} \\
\includegraphics[width=0.45\textwidth]{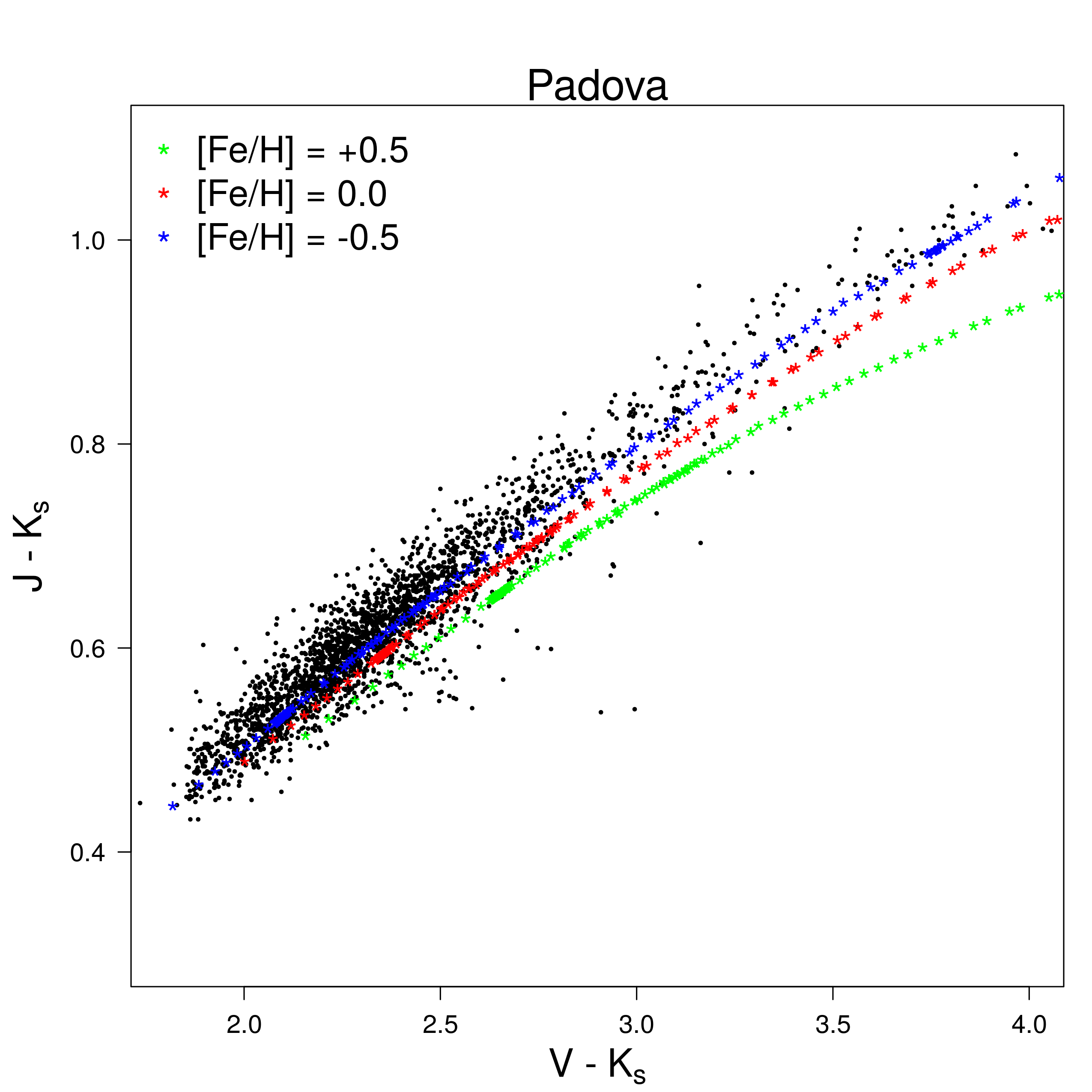} &
\includegraphics[width=0.45\textwidth]{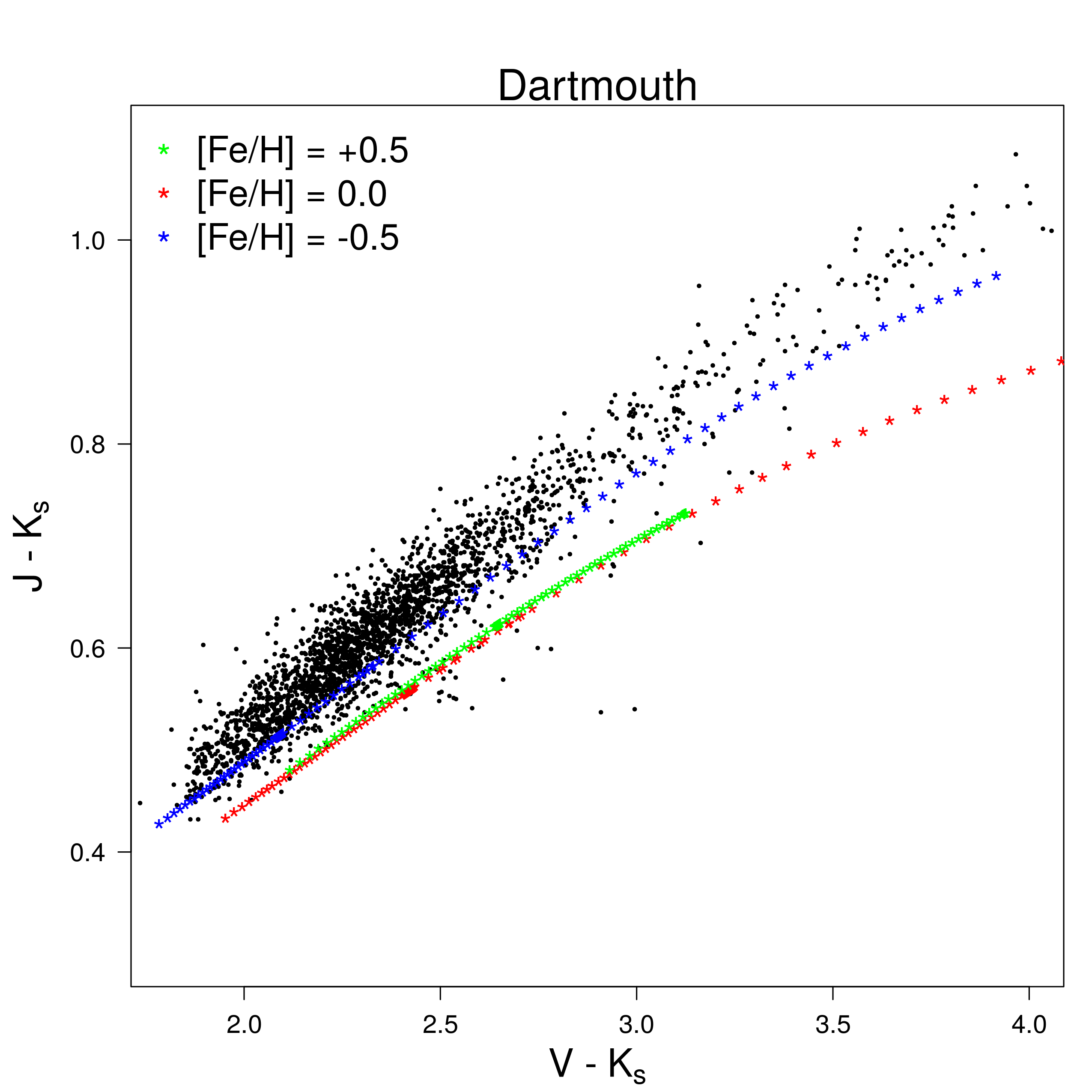} 
\end{tabular}
\caption{$B-V$ vs $V-I$ and $J-\ks$ vs $V-\ks$ colour-colour diagrams of RC stars (sample described in Section \ref{sect:data}). The median metallicity of the sample is about -0.2. \textit{Padova} Parsec 2.7 (left) and \textit{Dartmouth} (right) isochrones with a median age of 2Gyr are overplotted for three different metallicities: (green) $\feh = +0.5$, (red) $\feh = 0.0$, (blue) $\feh = -0.5$. Only the Red Giant Branch and the Early Assymptotic Giant Branch are shown}
\label{rcgap}
\end{figure*}

By comparing the observations with isochrones we can directly constrain stellar parameters such as ages and metallicities. However, we found that at the level of red giant stars, atmosphere models and observations do not fit: there is a \textit{gap} between them no matter the photometric bands nor the atmosphere models used. As an example, we show this issue in Fig. \ref{rcgap} for the $B-V$ vs $V-I$ and $J-\ks$ vs $V-\ks$ colour-colour diagrams of some RC stars, and for both \textit{Padova} (Parsec 2.7) and \textit{Dartmouth} isochrones, who use ATLAS and Phoenix atmosphere models, respectively. A more exhaustive work on the important effects on the choice of atmosphere models and other parameters may be found in \cite{aringer_synthetic_2016}. We have checked that a unique shift is not enough to correct this \textit{gap} because the slope is also different. We have also checked the influence of filter modelling. Nevertheless, it seems that it is most probably an issue of atmosphere models.

Therefore there are two aspects that led us to develop the purely empirical calibrations that we present in this work: 1) the need of a photometric calibration totally independent of models, and 2) the fact that there is no on-board Gaia calibrated filter profile (instrumental response) available for the GDR1, thus a colour-colour calibration was automatically needed. \cite{jordi_gaia_2010} already predicted some colour relationships based on theoretical spectra and the nominal Gaia passbands (calibrated before launch), but the effective filters actually slightly differ \citep{vanleuween_gdr1phot_2017}. Therefore, there is an special interest in using colour-colour empirical calibrations instead.

In this work we present the first metallicity-dependent empirical colour-colour (hereafter \textit{CC}), effective temperature ($\teff$-colour and colour-$\teff$, hereafter $\teff C$ and $C \teff$ respectively) and absolute magnitude ($M_G$ and $M_K$) calibrations for solar neighbourhood RC stars using the Gaia $G$ magnitude.

The paper is organized as follows. In Sect. \ref{sect:data} we describe the sample selection, the adopted constraints and how the interstellar extinction has been handled. The method developed to calibrate all the \textit{CC} and $\teff$ relations is explained in Sect. \ref{sect:method}. The calibrations obtained are presented in Sect. \ref{sect:res}. In Sect. \ref{sect:absmag} we detail the RC absolute magnitude calibration. And finally in Sect. \ref{sect:applications} we present the un-reddened TGAS HR diagram and compare its RC to the Padova isochrones.


\section{Sample selection} \label{sect:data}

Different samples were constructed using TGAS data for the colour-colour and the effective temperature calibrations. To ensure their quality we considered the following constraints.

\subsection{Interstellar extinction}

One of the main issues with \textit{CC}, $\teff C$ and $C \teff$ calibrations for giants is the extinction handling. To select low extinction stars, we use here the  most up-to-date 3D local extinction map of \cite{lallement_maps_2014}, \cite{capitanio_3dmap_2017}, together with the 2D \cite{schlegel_maps_1998} map for stars for which the distance go beyond the 3D map borders. We scaled the \cite{schlegel_maps_1998} map values by 0.884 according to \cite{schlafly_reddeningSDSS_2011} and in agreement with the \cite{capitanio_3dmap_2017} $E(B-V)$ scale. We fixed a maximum threshold of 0.01 in $E(B-V)$ (i.e. 0.03 in $A_0$) for a maximum distance corresponding to a parallax $\varpi - \sigma_\varpi$. Such a selection of low extinction stars should lead to more robust results than a derredening that would be dependent not only on an extinction map but also on an extinction law, and could lead to either over or under correction of the extinction.

\subsection{Red Giants selection}
To select solar neighbourhood red giant stars we considered the following two criteria:

\begin{align}
\gk > 1.6
\end{align}

\begin{align}
m_G + 5 + 5 \; \log_{10} \left(\frac{\varpi + 2.32 \; \sigma_{\varpi}}{1000} \right)  < 4.0 \label{eq:MGlim}
\end{align}

\begin{figure}
\centering
\begin{tabular}{c}
\includegraphics[width=0.45\textwidth]{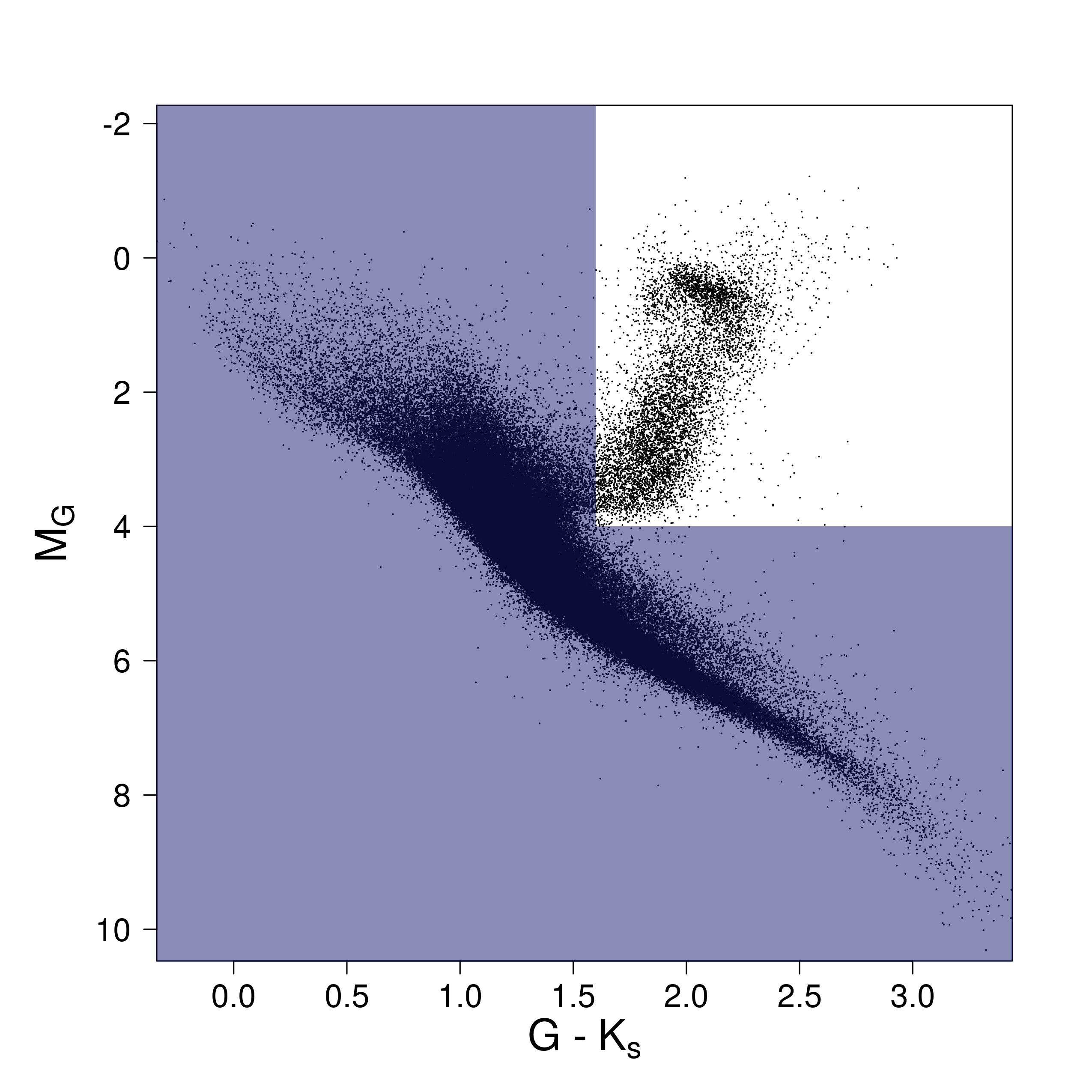}
\end{tabular}
\caption{TGAS HR diagram with parallax precision $\leq 10\%$, $\sigma_G < 0.01$, $E_{B-V} < $0.015 and 2MASS $JK$-bands high photometric quality (data available in Table \ref{tab:hrtgas}, Appendix \ref{annex:hrtgas}). The non-shaded region corresponds to the Red Giants selection, with $\gk > 1.6$ and $M_G < 4.0$}
\label{rcregion}
\end{figure}

The factor 2.32 on the parallax error corresponds to the 99th percentile of the parallax probability density function. Fig. \ref{rcregion} shows the selected region on the HR diagram. The data used to construct this HR diagram is described in Appendix \ref{annex:hrtgas}, Table \ref{tab:hrtgas}. See Sect. \ref{sect:applications} for more details on the RC region of this diagram. 

We extended the parallax criteria to cover the full red giant branch so that our calibrations have a larger interval of applicability than just the Red Clump. We checked that this large magnitude interval did not have any significant impact on the calibration. The fit is on the opposite very sensitive to red dwarf stars contaminants. Indeed the slope of giants and dwarfs in colour-colour distributions changes gradually as the stars are cooler \citep[e.g][]{bessellBrett_jhksystem_1988}. A selection based on spectroscopic surface gravity ($2.5 < \log g < 3.5$) was tested and discarded due to the non-negligible percentage of giants/dwarfs misidentification in some surveys (e.g. we found $\sim 2\%$ misidentified RAVE stars when selecting those matching Appendix \ref{annex:hrtgas} criteria and supposed to be inside the non-shaded region of Fig. \ref{rcregion}). The chosen parallax criteria allows us to guarantee there is no dwarfs' contamination in our sample.

\subsection{Photometric data} \label{sect:photdata}

Our calibrations aim to cover all major visual and infrared bands. To achieve this we selected only those DR1 stars which have photometric information (with uncertainties) from the following catalogues:

\subsubsection{GDR1}
$G$ band with uncertainties lower than 0.01 mag. An error of 10 mmag was quadratically added to mitigate the impact of bright stars residual systematics, see \cite{arenou_cu9validationDR1_2017, evans_dr1photometry_2017}.

\subsubsection{Hipparcos}
$B$, $V$ and $\hp$ bands with uncertainties lower than 0.03 mag. We did not include the I band because of the low number ($\sim12$) of remaining stars when selecting those with $V-I$ direct measurements in the Cousin's system (field $H42 = A$), with measurements in the Johnson's system then converted to Cousin's (field $H42 = C$), and with measurements in the Kron-Eggen's system then converted to Cousin's (field $H42 = E$). For more details see \cite{perryman1997hipparcosbook}, Vol. 1, Sect. 1.3, Appendix 5.

\subsubsection{Tycho-2}
$B_\textrm{T}$ and $V_\textrm{T}$ bands \citep{hog_tycho2_2000} with uncertainties lower than 0.03 mag.

\subsubsection{2MASS}
$J$, $H$ and $\ks$ bands \citep{cutri_2mass_2003} from the crossmatched 2MASS-GDR1 catalogue (Marrese et al. 2017, submitted) with high photometric quality (i.e. flag q2M = A) and from \cite{laney_jhk_2012}. Only stars with uncertainties lower than 0.03 mag.

\subsubsection{APASS DR9}
$g$, $r$ and $i$ bands \citep{henden_apassdr9_2016} crossmatched with Gaia at 2\arcsec\ precision, and only stars with standard deviations obtained from more than one observation ($u_e = 0$ flag in the APASS catalogue) and uncertainties lower than 0.03 mag. Duplicates were removed by keeping the source with the largest number of photometric bands provided in APASS.

\subsubsection{WISE}
$W1$, $W2$, $W3$ and $W4$ bands \citep{wright_wise_2010} from the crossmatched WISE-GDR1 catalogue (Marrese et al. 2017, submitted) with uncertainties lower than 0.05 mag, high photometric quality (i.e. flag qph = A), low probability of being true variables (i.e. flag var < 7), a source shape consistent with a point-source (i.e. flag ex = 0) and showing no contamination from artifacts (i.e. flag ccf = 0). According to \cite{cotten_comprehensiveNIR_2016}, for $W2$ we also removed stars brighter than 7 mag, because they are saturated.

\subsection{Binarity and Multiplicity}

We removed all stars flagged as binaries and belonging to multiple systems. To do so we took into account the specific flags in the \textit{Hipparcos} catalogue as well as the last updated information from:

	\begin{itemize}
		\item \textit{9th Catalogue of Spectroscopic Binary Orbits} \citep[SB9,][]{pourbaix_sb9first_2004, pourbaix_sb9_2009}
		\vspace{0.1cm}
		
		\item \textit{The Tycho Double Star Catalogue} \citep[TDSC,][]{fabricius_tdsc_2002}
		\vspace{0.1cm}	
		
		\item Simbad database (stars with flag "**")
		\vspace{0.1cm}	
		
		\item We also considered only stars for which the proper motions from Hipparcos are consistent with those of Tycho-2 (rejection p-value: 0.001). According to a specific test carried out in the framework of the Gaia data validation team \citep{arenou_cu9validationDR1_2017} most of the stars for which the proper motions are not consistent between both catalogues, are expected to be long period binaries not detected in Hipparcos, and for which the longer time baseline of Tycho-2 could have provided a more accurate value.

	\end{itemize}

\subsection{Metallicity}

We selected stars with metallicity information from different sources, since there were not enough stars when using just one reference. We expect the differences between all the measurements to increase the dispersion of the residuals and to decrease the dependence of the calibrations with metallicity. Noting in brackets the percentage of stars found and used in this work \footnote{Note that some references used in our compilation could lead to no star in the final sample, i.e. 0$\%$, due to the various quality cuts}, our established priority order is: \cite{morel_atmospheric_2014} [0\%], \cite{thygesen_atmospheric_2012} [0\%], \cite{bruntt_accurate_2012} [0.04\%], \cite{maldonado_abundances_2016} [0.6\%], \cite{alves_determination_2015} [1.4\%], \cite{jofre_stellar_2015} [0.4\%], \cite{bensby_abundances_2014} [0.4\%], \cite{dasilva_abundances_2015} [0.04\%], \cite{mortier_new_2013} [0.04\%], \cite{adibekyan_fgk_2012} [0.3\%], APOGEE DR13 \citep{albareti_apoDR13_2016} [9.9\%], GALAH \citep{martell_galah_2017} [0.4\%], \cite{ramirez_abundancesSunsiblings_2014} [0\%], \cite{ramirez_params_2014} [0\%], \cite{ramirez_abundances_2013} [0.04\%], \cite{zielinski_giants_2012} [0.2\%], \cite{puzeras_high-resolution_2010} [0.04\%], \cite{takeda_abundances_2008} [0.08\%], \cite{valentini_spectroscopic_2010} [0.6\%], \cite{saguner_spectroscopic_2011} [0\%], RAVE DR5 \citep{kunder_raveDR5_2017} [67.6\%], LAMOST DR2 \citep{luo_lamostDR2cat_2016} [13.5\%], AMBRE DR1 \citep{depascale_ambre_2014} [1.0\%], \cite{luck_abundances_2015} [0.7\%], PASTEL \citep{soubiran_pastelcat_2016} [2.8\%].  

The final sample contained 2334 stars when considering the extinction, red giants selection, multiplicity, metallicity and the photometric constraints on the $G$ and $\ks$ bands. Subsamples were then generated for each colour-colour relation depending on the other photometric bands used (see later in Table \ref{ccempirical} the final sizes for every fit).

\subsection{Effective Temperature} \label{sect:teffdata}

For the $\teff$ calibrations the largest homogeneous sample filling all the above criteria is the 13th release (DR13) of the APOGEE survey \citep{holtzman_apogee_2015, garciaperez_apogee_2016, albareti_apoDR13_2016}. To increase the sample size, we also include stars not in TGAS with APOGEE $\log g < 3.2$, using therefore only the \cite{schlegel_maps_1998} map to apply our low extinction criteria. The weighted mean of the parameters is computed for the duplicated sources. The cross-match with Gaia is done through the 2MASS cross-match (Marrese et al. 2017, submitted) with an angular distance $<$ 1\arcsec\ . The final sample contained 530 stars.

The SDSS Collaboration discuss a systematic offset \footnote{\url{http://www.sdss.org/dr13/irspec/parameters/\#QualityoftheASPCAPStellarParameters}} of their spectroscopic effective temperatures from photometrically-derived temperatures for metal-poor stars (by as much as 200-300 K for stars at $\feh \sim -2$). Consequently they provided a correction as a function of metallicity. We decided not to apply their suggested correction as it is based on comparison with photometric temperatures. We compared the APOGEE temperatures to the PASTEL ones and see metallicity correlations only for the most metal-poor stars ($\feh <$ -1.5). We tested that our calibrations did not change significantly when we removed the most metal-poor stars from our sample. 

\begin{figure}
\centering
\begin{tabular}{c}
\includegraphics[width=0.45\textwidth]{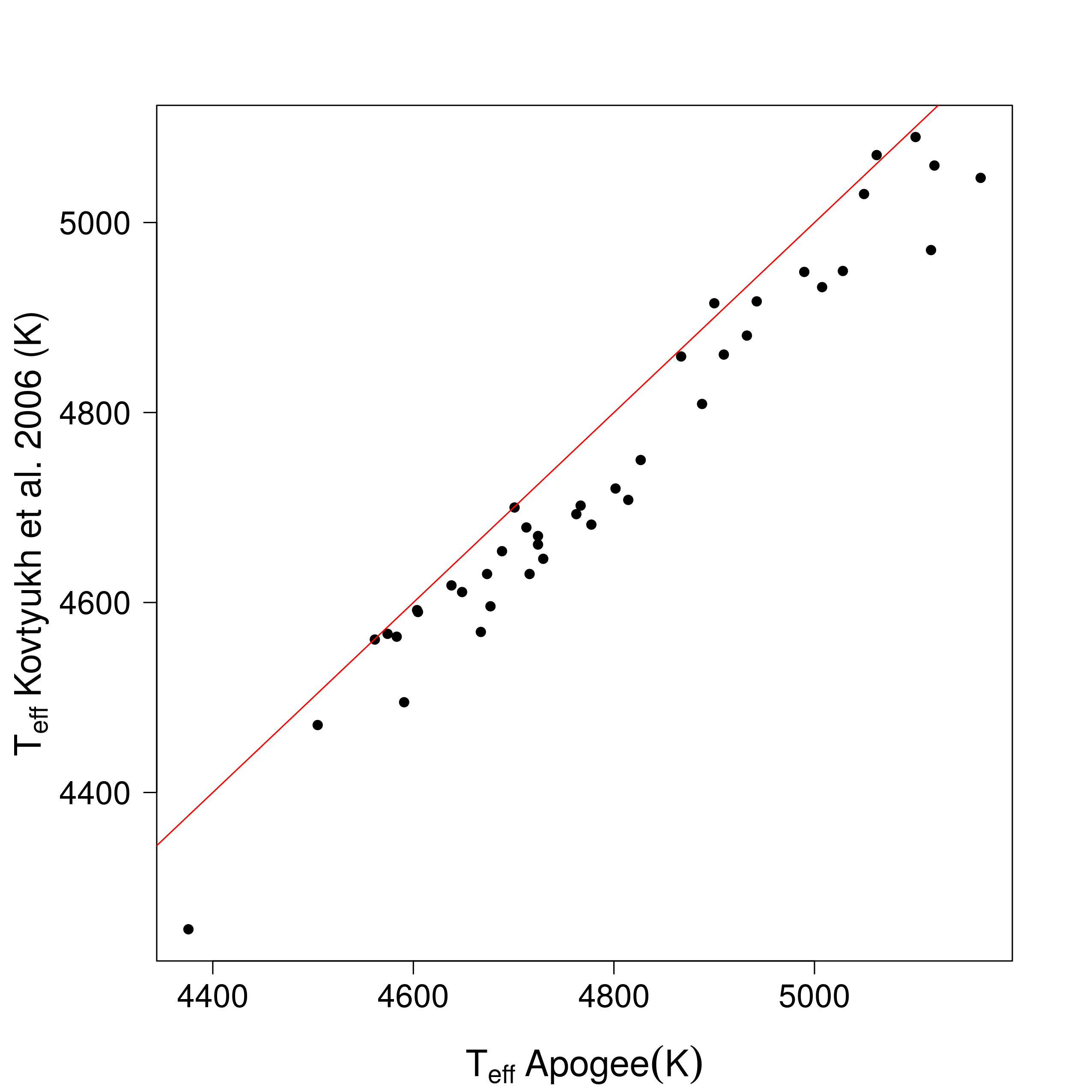}
\end{tabular}
\caption{Comparison of the spectroscopic effective temperatures of the 41 stars in common between APOGEE and \cite{kovtyukh_teffs_2006}}
\label{kovapo}
\end{figure}

For metal-rich stars, we compared 41 giant stars (within $-0.4 < \feh < 0.2$) from \cite{kovtyukh_teffs_2006} in common with APOGEE. They got a very good internal precision of 5-20K (zero-point difference expected to be smaller than 50K). As shown in Fig. \ref{kovapo} we find a difference of about 50K with respect to APOGEE, with no correlation with $\feh$, and a dispersion in agreement with the precisions provided in both catalogues.

\section{Calibration method} \label{sect:method}

To derive accurate photometric relations, we implemented a Monte Carlo Markov Chain (MCMC) method which allows us to account and deal with the uncertainties of both the predictor and response variables in a robust way.

We provide all the calibrations with respect to the $\gk$ colour. Those photometric bands will be widely used thanks to the all-sky and high uniformity properties of the Gaia and 2MASS catalogues. Thus, in this work we provide the following calibrations:

\begin{center}
\begin{equation}
\begin{aligned}
  \textrm{Colour} &= \textrm{f}(\textrm{$\gk$}, \feh) \\  
  \textrm{$\tnorm$} &= \textrm{f}(\textrm{$\gk$}, \feh) \\
  \textrm{$\gk$} &= \textrm{f}(\textrm{$\tnorm$}, \feh)
\end{aligned} \label{eq:functions}
\end{equation}
\end{center}

where \textit{Colour} includes all possible combinations of the photometric bands considered in this work (Section \ref{sect:photdata}), and $\hat{T} = T_\mathrm{eff}/5040$ is the normalised effective temperature.

\subsection{Polynomial models}

The general fitting formula adopted is:
	
	\begin{equation}
		Y = a_0 + a_1\;X + a_2\;X^2 + a_3\;\feh + a_4\;\feh^2 + a_5\;X\;\feh \label{eq:cc}
	\end{equation}

	a second order polynomial \footnote{Upper degrees were tested, but discarded by an Analysis of Variance test (ANOVA), meaning that simpler models were good enough to describe the data} where, following Eq. \ref{eq:functions}, $X$ is either the $\gk$ or the normalized effective temperature $\tnorm$, $Y$ is (for \textit{CC}) a given colour to be calibrated or (for $\teff$ relations) either $\gk$ or $\tnorm$, and $a_i$ are the coefficients to be estimated. In order to provide the most accurate fit for each relation, the process (see Sect. \ref{sect:dic}) penalises by the complex terms so that, in the end, seven different models may be tested for every relation (Model 7 being the more complex one):

$ \textrm{Model 1}: Y = a_0 + a_1\;X$

$ \textrm{Model 2}: Y = a_0 + a_1\;X  + a_2\;X^2$

$ \textrm{Model 3}: Y = a_0 + a_1\;X + a_3\;\feh$

$ \textrm{Model 4}: Y = a_0 + a_1\;X + a_2\;X^2 + a_3\;\feh$

$ \textrm{Model 5}: Y = a_0 + a_1\;X + a_3\;\feh + a_4\;\feh^2$

$ \textrm{Model 6}: Y = a_0 + a_1\;X + a_2\;X^2 + a_3\;\feh + a_4\;\feh^2$

$ \textrm{Model 7}: Y = a_0 + a_1\;X + a_2\;X^2 + a_3\;\feh + a_4\;\feh^2 + a_5\;X\;\feh$

Input uncertainties from all variables are taken into account in the model.

\subsection{MCMC}

A Monte Carlo Markov Chain was run for every model tested, with 10 chains and 10000 iterations for each. We used the \texttt{runjags} \footnote{\url{https://cran.r-project.org/web/packages/runjags/runjags.pdf}} library from the R programme language. An uninformative prior was set through a normal distribution with zero mean and standard deviation 10. Further, we also set an initial value for every coefficient. That is, we used the output coefficients obtained for each model through a multiple linear regression (simpler method which does not take uncertainties into account, but allows to obtain approximated values). The MCMC fit is run on the standardized variables to improve the efficiency of MCMC sampling (reducing the autocorrelation in the chains). Chain convergence is checked with the Gelman and Rubin's convergence diagnostic.

\subsection{Best model selection: DIC} \label{sect:dic}

The model selection was done through a process of penalisation by the complex terms. To do so we took advantage of the Deviance Information Criterion (DIC) \citep{plummer_dic_2008}, and we tested the models by pairs: a given complex model is compared to the next simpler model (e.g. we remove the highest-order interactions, starting with the cross-term $X*\feh$). 

The method continuously determined the next pair of models to be compared, run the MCMC for each and checked their DIC. When the DIC difference was significantly negative at 1 $\sigma$ (i.e. $\Delta DIC + \sigma_{\Delta DIC} < 0$) the complex model was kept, else the next pair was tested.

\subsection{Outliers}

Once the best model determined, the method checked whether there were calibrated stars out of 3$\sigma$ from the model. If so, the furthest one was removed and the complete process was run again. Outliers were eliminated one by one to ensure that the further one was not causing a deviation in the model that led to consider other stars as "false outliers".

\section{Calibration results} \label{sect:res}

\subsection{Colour-Colour relations} \label{sect:ccresults}

\begin{figure*}
\centering
\begin{tabular}{cc}
\includegraphics[width=0.45\textwidth]{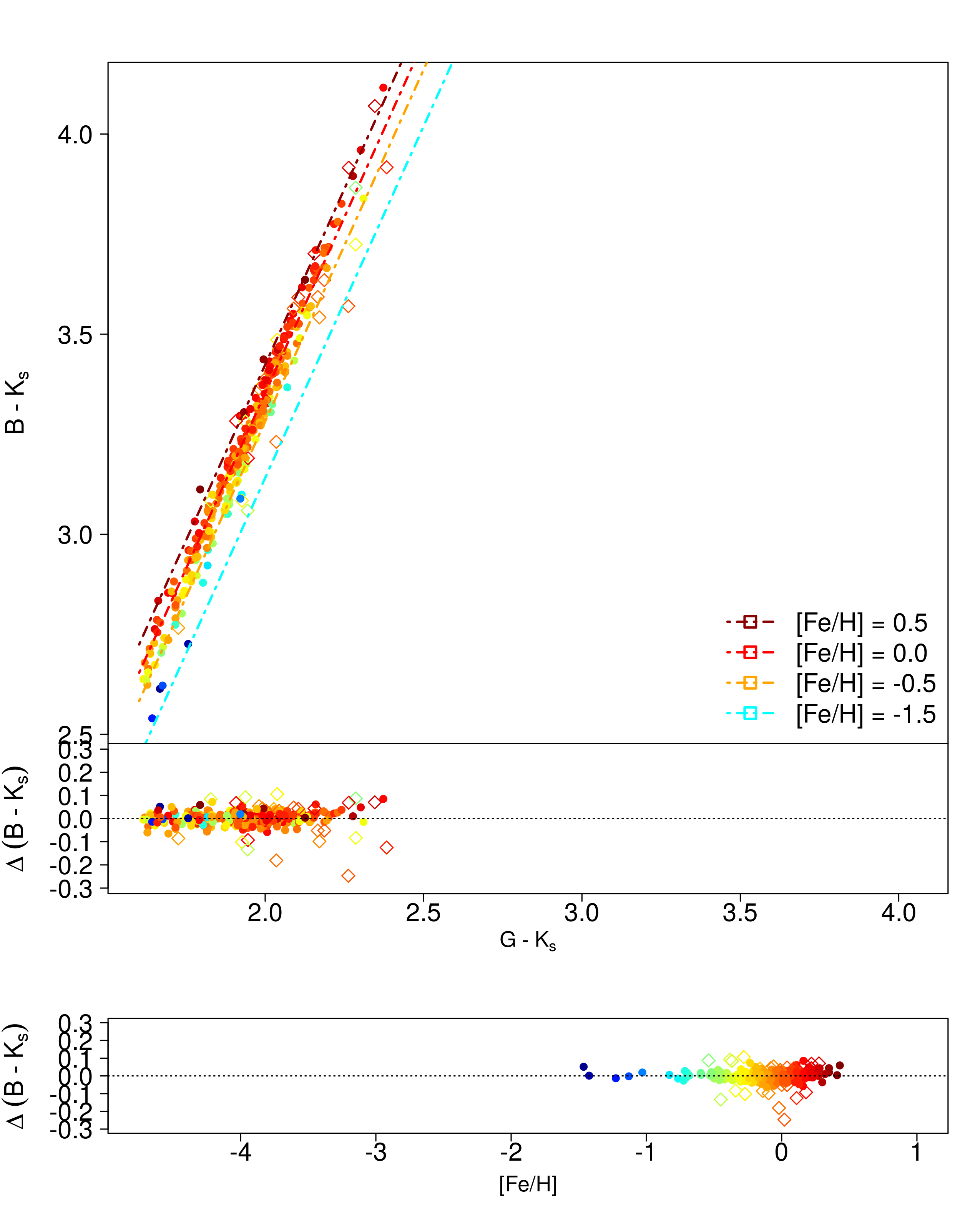} &
\includegraphics[width=0.45\textwidth]{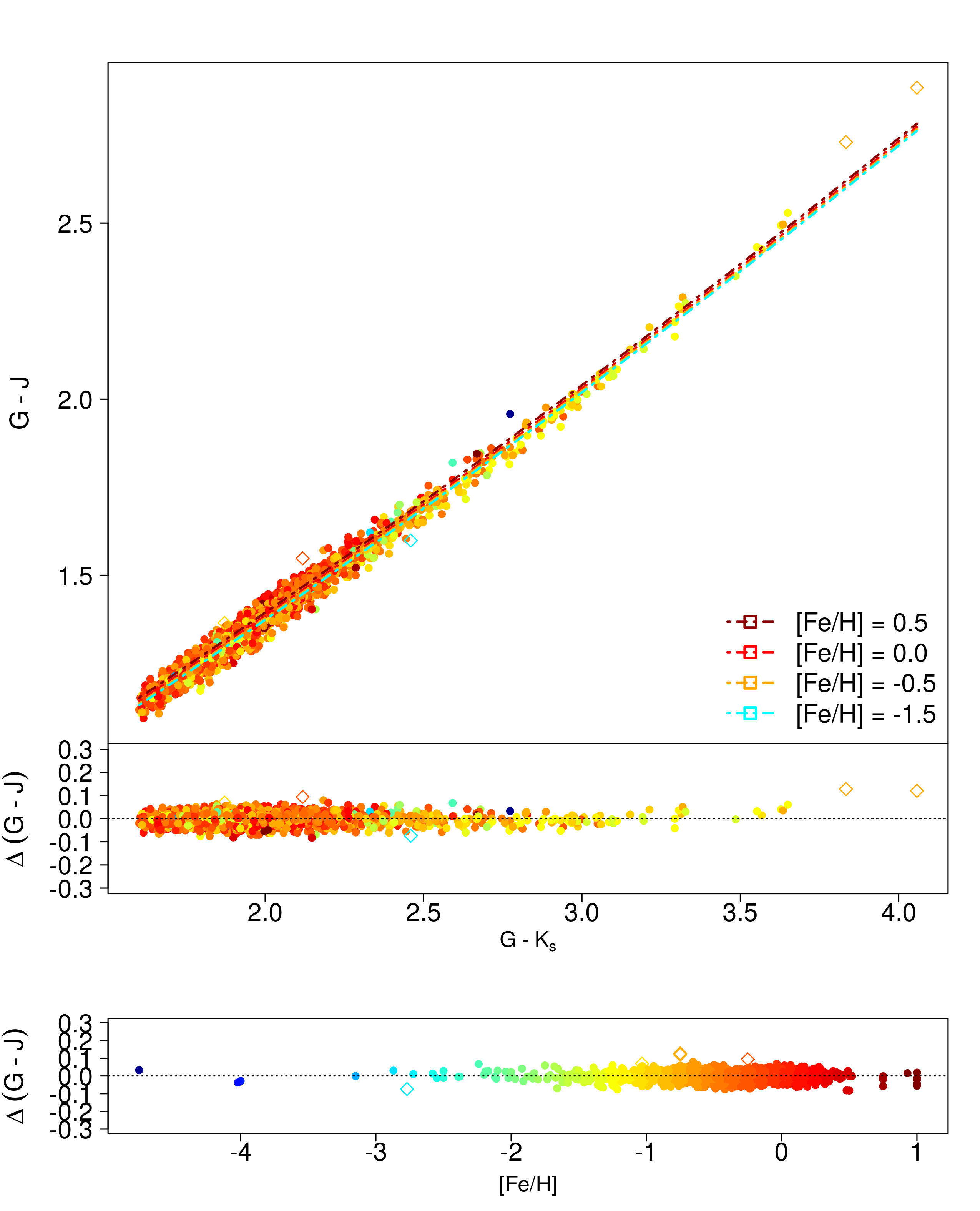} \\
\includegraphics[width=0.45\textwidth]{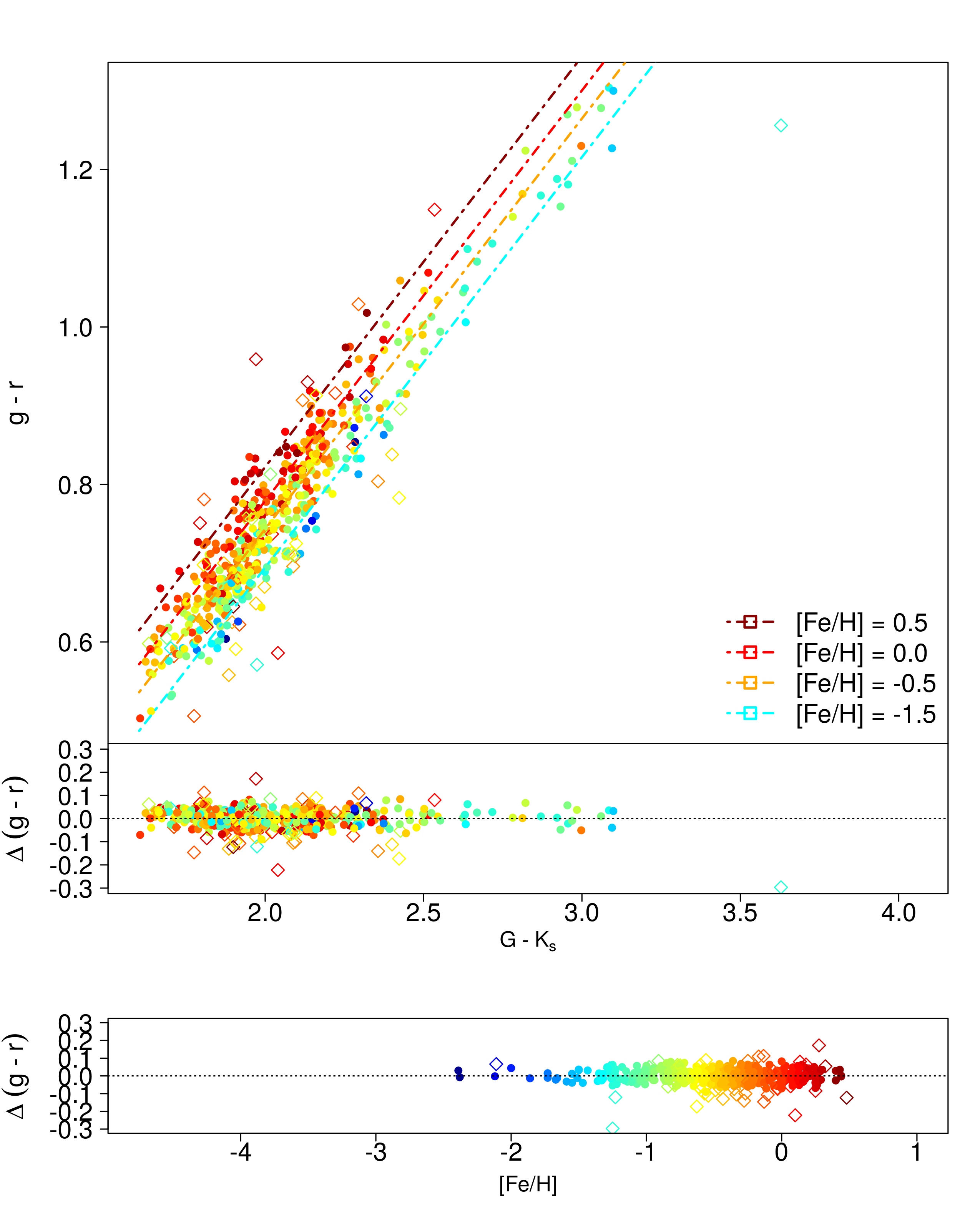} &
\includegraphics[width=0.45\textwidth]{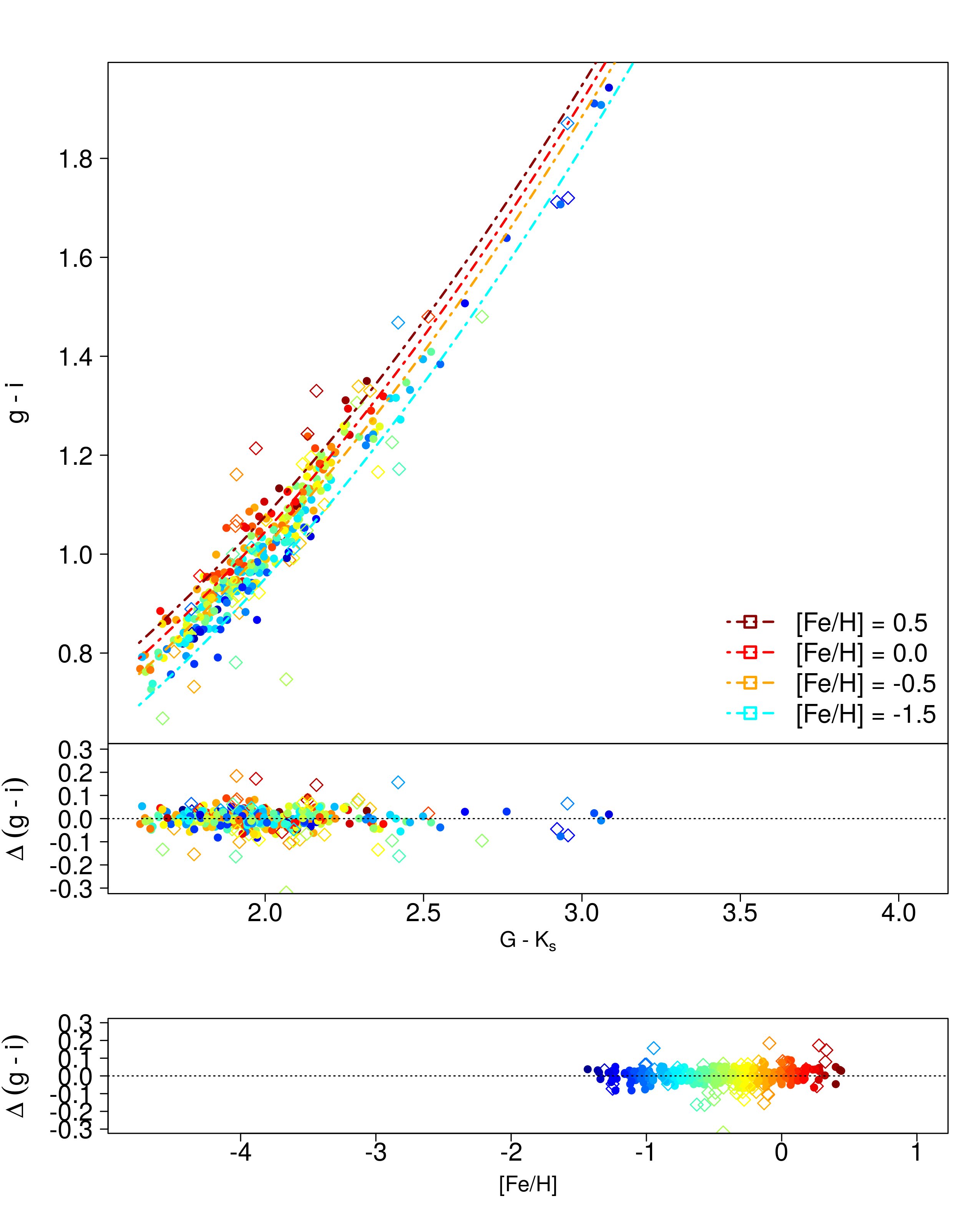} \\
\end{tabular}
\caption{Empirical colour vs $\gk$ calibrations for $B-\ks$, $G-J$, $g-r$ and $g-i$ colour indices. The dash-dotted lines correspond to our calibration at fixed metallicities (see legend). The colour of the points varies as a function of the metallicity. Diamonds correspond to the outliers removed at 3$\sigma$ during the MCMC process. At the bottom of each calibration plot: residuals of the fit as a function of $\gk$ and $\feh$. All plots are scaled to the same $\gk$ colour and metallicity intervals}
\label{ccempirical}
\end{figure*}

Table \ref{tab:cccoeffs} gives the coefficients for each of the 19 colour vs $\gk$ fit, together with the $\gk$ and metallicity ranges of applicability (defined by the maximum and minimum values of each individual sample), as well as the number (N) of stars used after the 3$\sigma$ clipping, the percentage of outliers removed and the final root mean square deviation (RMS). Fig. \ref{ccempirical} shows the colour vs $\gk$ relations obtained for four of the twenty colour indices with the residuals of the fit as a function of the colour itself and of the metallicity. The scatter obtained in the residuals is very small ($\sim \pm 0.03$ globally).

\begin{table*}
\caption{Coefficients and range of applicability of colour vs $\gk$ relations: $Y = a_0 + a_1\;(\gk) + a_2\;(\gk)^2 + a_3\;\feh + a_4\;\feh^2 + a_5\;(\gk) \;\feh$}\label{tab:cccoeffs}
\begin{center}
\resizebox{\textwidth}{!}{\begin{tabular}{l|c|c|r|r|r|r|r|r|c|c|r} \hline\hline
\centering \bf Colour & \textbf{$\gk$ range} & \textbf{$\feh$ range} &  \multicolumn{1}{c|}{$\bm{a_0}$} & \multicolumn{1}{c|}{$\bm{a_1}$} & \multicolumn{1}{c|}{$\bm{a_2}$}  & \multicolumn{1}{c|}{$\bm{a_3}$} & \multicolumn{1}{c|}{$\bm{a_4}$}  & \multicolumn{1}{c|}{$\bm{a_5}$}  & \multicolumn{1}{c|}{\bf RMS} & \multicolumn{1}{c|}{$\bm{\%_{\rm outliers}}$ } & \multicolumn{1}{c}{\bf N}   \\\hline 
 ${\textrm{B - G}}$  & [1.6, 2.4] & [-1.4, 0.4] & 0.583 $\pm$ 0.180 & -0.046 $\pm$ 0.187 & 0.215 $\pm$ 0.049 & 0.144 $\pm$ 0.006 & \multicolumn{1}{c|}{-} & \multicolumn{1}{c|}{-} & 0.02 & 17.9 & 230\\ 
   ${\textrm{B - V}}$  & [1.6, 2.4] & [-1.4, 0.4] & -0.094 $\pm$ 0.017 & 0.552 $\pm$ 0.009 & \multicolumn{1}{c|}{-} & 0.129 $\pm$ 0.005 & \multicolumn{1}{c|}{-} & \multicolumn{1}{c|}{-} & 0.02 & 10.4 & 251 \\ 
   ${\textrm{B - J}}$  & [1.6, 2.4] & [-1.5, 0.4] & -0.117 $\pm$ 0.041 & 1.432 $\pm$ 0.021 & \multicolumn{1}{c|}{-} & 0.153 $\pm$ 0.011 & \multicolumn{1}{c|}{-} & \multicolumn{1}{c|}{-} & 0.03 & 12.9 & 176 \\ 
   ${\textrm{B -  $\ks$}}$  & [1.6, 2.4] & [-1.5, 0.4] & -0.161 $\pm$ 0.038 & 1.757 $\pm$ 0.020 & \multicolumn{1}{c|}{-} & 0.141 $\pm$ 0.011 & \multicolumn{1}{c|}{-} & \multicolumn{1}{c|}{-} & 0.02 & 9.3 & 254  \\ 
   ${\textrm{G  - $\hp$}}$  &  [1.6, 2.4] & [-1.5, 0.4] & 0.029 $\pm$ 0.009 & -0.270 $\pm$ 0.005 & \multicolumn{1}{c|}{-} & -0.023 $\pm$ 0.003 & \multicolumn{1}{c|}{-} & \multicolumn{1}{c|}{-} & 0.01 & 5.3 & 270 \\ 
   ${\textrm{G - V}}$  & [1.6, 2.4] & [-1.5, 0.4] & -0.286 $\pm$ 0.104 & 0.191 $\pm$ 0.107 & -0.110 $\pm$ 0.028 & -0.017 $\pm$ 0.003 & \multicolumn{1}{c|}{-} & \multicolumn{1}{c|}{-} & 0.01 & 3.9 & 274 \\ 
   ${\textrm{G} - \textrm{B}_{\mbox{\textrm{\scriptsize T}}}}$  & [1.6, 2.4] & [-1.4, 0.4] & -0.375 $\pm$ 0.257 & -0.194 $\pm$ 0.267 & -0.218 $\pm$ 0.069 & -0.201 $\pm$ 0.009 & \multicolumn{1}{c|}{-} & \multicolumn{1}{c|}{-} & 0.03 & 12.7 & 241  \\ 
   ${\textrm{G} - \textrm{V}_{\mbox{\textrm{\scriptsize T}}}}$  & [1.6, 2.4] & [-1.5, 0.4] & -0.261 $\pm$ 0.115 & 0.122 $\pm$ 0.119 & -0.109 $\pm$ 0.031 & -0.034 $\pm$ 0.006 & -0.016 $\pm$ 0.007 & \multicolumn{1}{c|}{-} & 0.01 & 3.5 & 272 \\ 
   ${\textrm{G - J}}$  & [1.6, 3.6] & [-4.8, 1.0] & 0.256 $\pm$ 0.021 & 0.510 $\pm$ 0.019 & 0.027 $\pm$ 0.004 & 0.016 $\pm$ 0.002 & 0.005 $\pm$ 0.001 & \multicolumn{1}{c|}{-} & 0.02 & 0.2 & 2178 \\ 
   ${\textrm{V - J}}$  & [1.6, 2.4] & [-1.5, 0.4] & -0.028 $\pm$ 0.026 & 0.880 $\pm$ 0.013 & \multicolumn{1}{c|}{-} & \multicolumn{1}{c|}{-} & \multicolumn{1}{c|}{-} & \multicolumn{1}{c|}{-} & 0.03 & 2.4 & 200 \\ 
   ${\textrm{V - $\ks$}}$  & [1.6, 2.4] & [-1.5, 0.4] & 0.326 $\pm$ 0.231 & 0.786 $\pm$ 0.237 & 0.112 $\pm$ 0.061 & 0.019 $\pm$ 0.008 & \multicolumn{1}{c|}{-} & \multicolumn{1}{c|}{-} & 0.01 & 2.1 & 279 \\ 
   ${\textrm{J - $\ks$}}$  & [1.6, 3.6] & [-4.8, 1.0] & -0.227 $\pm$ 0.024 & 0.466 $\pm$ 0.021 & -0.023 $\pm$ 0.005 & -0.016 $\pm$ 0.002 & -0.005 $\pm$ 0.001 & \multicolumn{1}{c|}{-} & 0.02 & 0.1 &  2180 \\ 
   $\textrm{B}_{\mbox{\textrm{\scriptsize T}}} -  \textrm{V}_{\mbox{\textrm{\scriptsize T}}}$  & [1.6, 2.4] & [-1.5, 0.4] & -0.247 $\pm$ 0.023 & 0.713 $\pm$ 0.012 & \multicolumn{1}{c|}{-} & 0.175 $\pm$ 0.007 & \multicolumn{1}{c|}{-} & \multicolumn{1}{c|}{-} & 0.03 & 8.0 & 254  \\ 
   ${\textrm{g - r}}$  & [1.6, 3.1] & [-2.4, 0.4] & -0.263 $\pm$ 0.010 & 0.521 $\pm$ 0.005 & \multicolumn{1}{c|}{-} & 0.079 $\pm$ 0.006 & 0.015 $\pm$ 0.004 & \multicolumn{1}{c|}{-} & 0.03 & 8.8 &  465\\ 
   ${\textrm{g - i}}$ & [1.6, 3.1] & [-1.4, 0.4] & 0.280 $\pm$ 0.084 & 0.057 $\pm$ 0.079 & 0.163 $\pm$ 0.018 & 0.063 $\pm$ 0.005 & \multicolumn{1}{c|}{-} & \multicolumn{1}{c|}{-} & 0.03 & 13.5 & 282 \\ 
   ${\textrm{r - i}}$  & [1.6, 3.1] & [-1.4, 0.4] & 0.236 $\pm$ 0.050 & -0.171 $\pm$ 0.047 & 0.095 $\pm$ 0.011 & \multicolumn{1}{c|}{-} & \multicolumn{1}{c|}{-} & \multicolumn{1}{c|}{-} & 0.02 & 2.2 & 364 \\
   ${\textrm{G - W1}}$  & [1.6, 3.2] & [-2.4, 0.5] & 0.099 $\pm$ 0.043 & 0.948 $\pm$ 0.040 & 0.019 $\pm$ 0.009 & 0.006 $\pm$ 0.004 & 0.007 $\pm$ 0.003 & \multicolumn{1}{c|}{-} & 0.03 & 0.4 & 1666 \\
   ${\textrm{W1 - W2}}$  & [1.6, 3.2] & [-2.4, 0.5] & 0.065 $\pm$ 0.039 & -0.051 $\pm$ 0.038 & -0.014 $\pm$ 0.009 & 0.049 $\pm$ 0.015 & 0.007 $\pm$ 0.002 & -0.028 $\pm$ 0.008 & 0.02 & 0.1 & 1657   \\
   ${\textrm{W2 - W3}}$  & [1.6, 3.2] & [-2.4, 0.5] & -0.228 $\pm$ 0.032 & 0.240 $\pm$ 0.029 & -0.038 $\pm$ 0.006 & \multicolumn{1}{c|}{-} & \multicolumn{1}{c|}{-} & \multicolumn{1}{c|}{-} & 0.03 & 0.1 & 1671 \\ 
  ${\textrm{H - W2}}$  & [1.6, 3.2] & [-2.4, 0.5] & 0.025 $\pm$ 0.008 & 0.032 $\pm$ 0.004 & \multicolumn{1}{c|}{-} & 0.009 $\pm$ 0.004 & 0.016 $\pm$ 0.003 & \multicolumn{1}{c|}{-} & 0.03 & 0.4 & 1137
  \\\hline
\end{tabular}}
\end{center}
\end{table*}

\subsection{Effective Temperature calibration}

Table \ref{tab:teffccoeffs} provides the coefficients of the fit for the $\tnorm$ vs $\gk$ and the $\gk$ vs $\tnorm$ calibrations. The colour, $\teff$ and metallicity ranges of applicability are specified, as well as the number (N) of stars used after the 3$\sigma$ clipping, the percentage of outliers removed and the final root mean square deviation.

\begin{figure}
\centering
\begin{tabular}{c}
\includegraphics[width=0.45\textwidth]{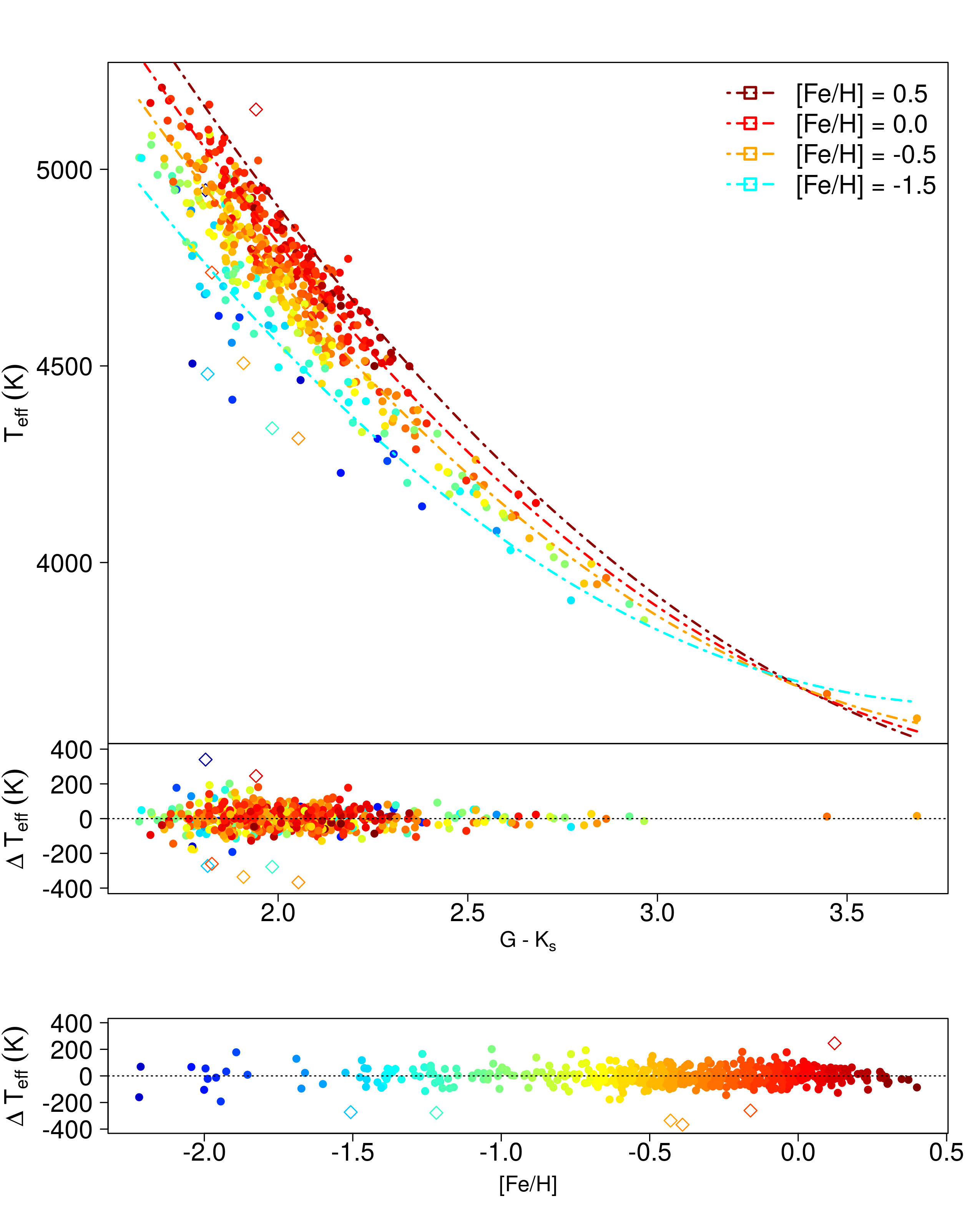}
\end{tabular}
\caption{Empirical normalised effective temperature versus $\gk$ calibration. The dash-dotted lines correspond to our calibration at fixed metallicities (see legend). The colour of the points varies as a function of the metallicity. Diamonds correspond to the outliers removed at 3$\sigma$ during the MCMC process. At the bottom: residuals of the fit as a function of $\gk$ and $\feh$}
\label{TeffCempirical}
\end{figure}

Figure \ref{TeffCempirical} shows the $\tnorm$ vs $\gk$ relation obtained with the residuals as a function of $\gk$ colour index and the metallicity. Since previous works in the literature use $\theta = 5040 / \teff$ instead of the $\tnorm$ considered here (e.g. \cite{ramirez_effective_2005}, \cite{gonzalez_hernandez_new_2009} or \cite{huang_empirical_2015}), we also computed the calibration by using $\theta$. We found both calibrations look similar except for the cool stars, for which we just have a few points. We may see how in this region the $\tnorm$ relations at different metallicities cross each other in an unrealistic way. This does not happen for the $\theta$ fit. However, after having statistically compared both the $\tnorm$ and $\theta$ calibrations, we chose to provide only the coefficients for $\tnorm$ vs $\gk$. Indeed, DIC is significantly lower for the $\tnorm$ fit. The dispersion obtained on the $\teff$ residuals is about 59K, consistent with the uncertainties of the APOGEE data used ($\sim 69$K).

\begin{table*}
\caption{Coefficients and range of applicability of the $\tnorm$ vs $\gk$ relation (top table) and of the [$\tnorm$, $\gk$] relation (bottom table): $Y = a_0 + a_1\;X + a_2\;X^2 + a_3\;\feh + a_4\;\feh^2 + a_5\;X \;\feh$. We remind that $\tnorm = \teff / 5040$. Note that the range of temperatures of the [$\tnorm$, $\gk$] calibration (second table) is given in $\teff$ (not $\tnorm$)}\label{tab:teffccoeffs}
\begin{center}
\resizebox{\textwidth}{!}{\begin{tabular}{c|c|c|r|r|r|r|r|r|c|c|r} \hline\hline
\multicolumn{1}{c|}{\textbf{$\teff$}} & \textbf{$\gk$ range} & \textbf{$\feh$ range} &  \multicolumn{1}{c|}{$\bm{a_0}$} & \multicolumn{1}{c|}{$\bm{a_1}$} & \multicolumn{1}{c|}{$\bm{a_2}$}  & \multicolumn{1}{c|}{$\bm{a_3}$} & \multicolumn{1}{c|}{$\bm{a_4}$}  & \multicolumn{1}{c|}{$\bm{a_5}$} & \multicolumn{1}{c|}{\bf RMS$_{[\teff (K)]}$} & \multicolumn{1}{c|}{$\bm{\%_{\rm outliers}}$ } & \multicolumn{1}{c}{\bf N} \\\hline 
$\tnorm$ & [1.6, 3.7] & [-2.2, 0.4] & 1.648 $\pm$ 0.027 & -0.455 $\pm$ 0.023 & 0.054 $\pm$ 0.005 & 0.088 $\pm$ 0.012 & 0.001 $\pm$ 0.002 & -0.026 $\pm$ 0.006 & 59 & 1.3 & 523  \\\hline
\multicolumn{11}{c}{} \\\hline 
\multicolumn{1}{c|}{\textbf{Colour}} & \textbf{$\teff$ range} (K) & \textbf{$\feh$ range} &  \multicolumn{1}{c|}{$\bm{a_0}$} & \multicolumn{1}{c|}{$\bm{a_1}$} & \multicolumn{1}{c|}{$\bm{a_2}$}  & \multicolumn{1}{c|}{$\bm{a_3}$} & \multicolumn{1}{c|}{$\bm{a_4}$}  & \multicolumn{1}{c|}{$\bm{a_5}$}  & \multicolumn{1}{c|}{\bf RMS$_{[\gk]}$} & \multicolumn{1}{c|}{$\bm{\%_{\rm outliers}}$ } & \multicolumn{1}{c}{\bf N} \\\hline 
$\gk$ & [3603.7, 5207.7] & [-2.2, 0.4] & 13.554 $\pm$ 0.478 & -20.429 $\pm$ 1.020 & 8.719 $\pm$ 0.545 & 0.143 $\pm$ 0.013 & -0.0002 $\pm$ 0.009 & \multicolumn{1}{c|}{-} & 0.05 & 1.3 & 523 \\\hline
\end{tabular}}
\end{center}
\end{table*}

\subsection{Comparison with other studies}  \label{sect:discussion}

In order to test the metallicity-dependent $\teff C$ calibration derived in the current work, we took advantage of the already existing effective temperature relations provided by some studies. The closest literature relations to our $\tnorm$ vs $\gk$ calibration are $\teff$ vs $V-\ks$. We therefore selected a sample of APOGEE stars, with photometry information on the $G$, $V$ and $\ks$ bands, and satisfying the quality criteria specified in Section \ref{sect:data}. This gave us 179 stars for the test. Their effective temperatures were calculated by using our $\tnorm$ vs $\gk$ relation and through the $\teff$ vs $V-\ks$ relations from three different studies:

\begin{itemize}

\item \cite{ramirez_effective_2005}: calibrations for main-sequence and giant stars based on temperatures derived with the infrared flux method (IRFM). They are valid within a range of temperatures and metallicities of 4000 K - 7000 K and -3.5 $\leq \feh \leq$ 0.4, respectively, and spectral types F0 to K5. The calibrations were done using a sample of more than 100 stars with known $UBV$, $uvby$, Vilnius, Geneva, $RI$ (Cousins), DDO, Hipparcos-Tycho and 2MASS photometric bands.

\vspace{0.3cm}

\item \cite{gonzalez_hernandez_new_2009}: also with the IRFM, they derived a new effective temperature scale for FGK stars, by using the 2MASS catalogue and theoretical fluxes computed from ATLAS models. Their $\teff$-colour calibrations obtained with these temperatures are especially meant to be good for metal-poor stars. The calibrations were done using Johnson-Cousins $BV$($RI$), the 2MASS $JH\ks$ photometric bands and the Strömgren $b-y$ colour index 

\vspace{0.3cm}

\item \cite{huang_empirical_2015}: calibrations for dwarfs and giants based on a collection from the literature of about two hundred nearby stars (including 54 giants) with direct interferometry effective temperature measurements. Their giant's calibrations are valid for an effective temperature range of 3100 K - 5700 K and spectral types K5 to G5. The calibrations were done using Johnson $UBVRIJHK$, the Cousins $I_\textrm{C}R_\textrm{C}$, the 2MASS $JH\ks$ and the SDSS gr photometric bands	

\end{itemize}

\begin{figure}
\centering
\begin{tabular}{c}
\includegraphics[width=0.47\textwidth]{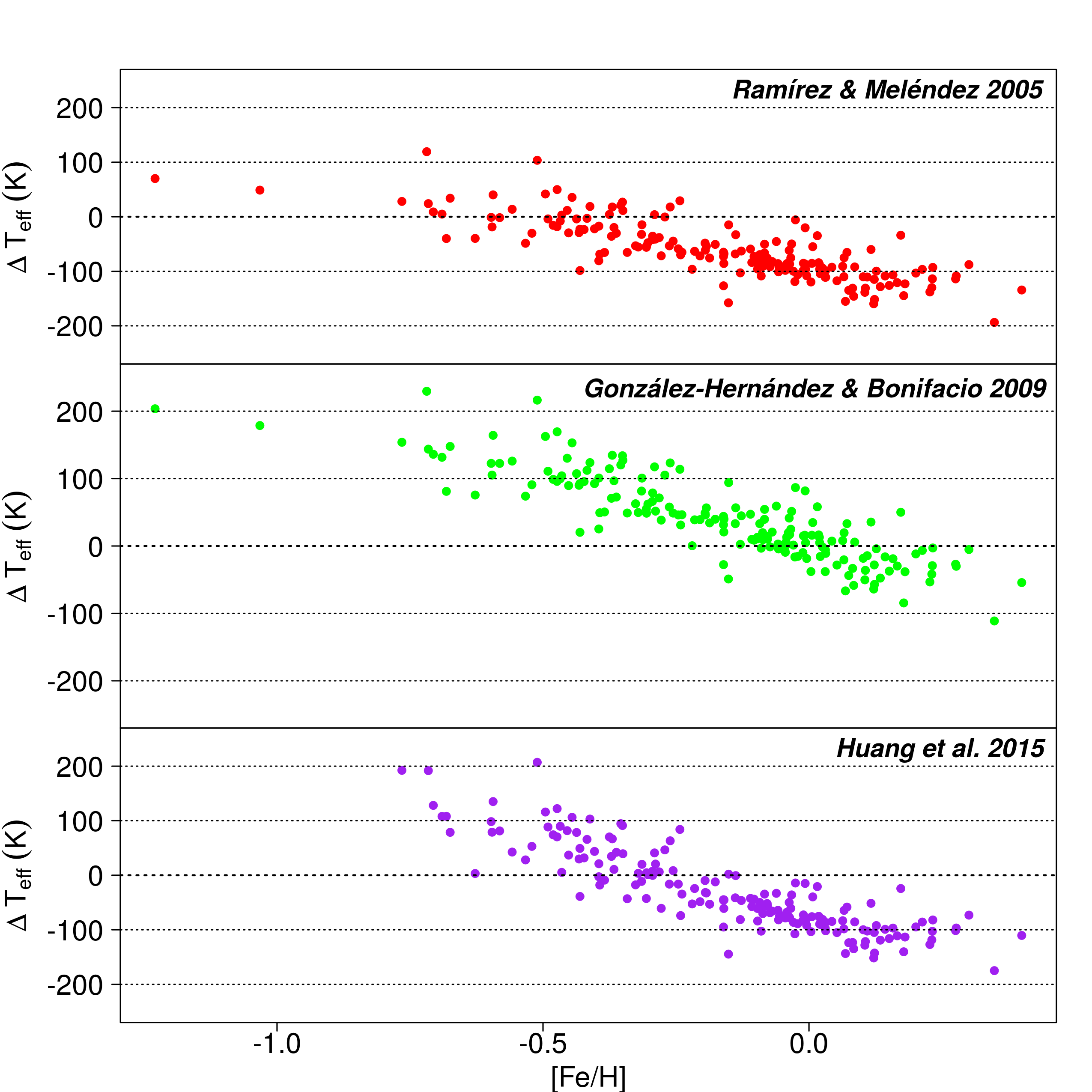}
\end{tabular}
\caption{Residuals of the effective temperatures obtained through the $\teff$ vs $V-\ks$ calibrations of \cite{ramirez_effective_2005} (top panel, red), \cite{gonzalez_hernandez_new_2009} (middle panel, green) and \cite{huang_empirical_2015} (bottom panel, purple), with respect to the values derived with the $\tnorm$ vs $\gk$ fit of this work, for a sample of 179 APOGEE stars with high photometric quality and low interstellar extinction.}
\label{teffcompar}
\end{figure}

Fig. \ref{teffcompar} shows the residuals of the effective temperatures obtained with these various $\teff$ vs $V-\ks$ literature relations with respect to the $\teff$ derived through our $\tnorm$ vs $\gk$ fit. The differences stay mostly within $\pm$100K (mainly explained by the various sources of effective temperatures together with the different treatment of the interstellar extinction in each case) but with a strong correlation with metallicity. This correlation is stronger than the one documented on the APOGEE DR13 release (see the corresponding discussion in Section 2.6). Applying the suggested correction still leads to a significant correlation with the residuals of the \cite{gonzalez_hernandez_new_2009} temperatures, although smaller. For users interested in working in the \cite{gonzalez_hernandez_new_2009} [GHB09] temperature scale, we found that, within the range of metallicity tested here,  a simple linear relation allows the transformation : ${\teff}_{(\textrm{GHB09})} =  5040 \; \tnorm + 7 -200 \; \feh$.

\section{The RC absolute magnitudes}   \label{sect:absmag} 

To calibrate the absolute magnitudes of the RC, we selected a different sample of stars. Indeed, in order to avoid contamination by the Secondary Red Clump (see next Sect. \ref{sect:applications}), we made use only of stars within 1.93 $< \gk <$ 2.3 and for which $M_{\ks}$ is brighter than -0.5 (similar as in Eq. \ref{eq:MGlim}). As in Sect. \ref{sect:data}, we kept only low extinction stars (i.e. $E(B-V)_{\rm max} < 0.01$), with $\sigma_G < 0.01$ and high photometric quality on the $K$ band. The sample contains 2482 stars. For each band we then applied the same photometric constraints as for the \textit{CC} and $\teff C$ calibrations, previously specified in Sect. \ref{sect:photdata}.

Considering the strong contamination of the RC by the RGB bump and the variation of both the RC and the RGB bump with colour, we did not estimate the RC absolute magnitude through a Gaussian fit to the magnitude distribution but through the mode of the distribution. The mode estimate is also less sensitive to the sample selection function. To model the colour dependency we looked for the maximum of $Q(\alpha,\beta)$, a kernel based distribution function of the residuals $M_{\lambda}-(\alpha (\gk) + \beta)$, with $M_{\lambda}$ the absolute magnitude of each particular band.

\begin{equation}
\max_{(\alpha,\beta)} Q(\alpha,\beta) = \sum_{i=1}^N \phi(\alpha + \beta \; (\gk-2.1) - M_{\lambda})
\end{equation}

where the constant 2.1, the median of $\gk$ of the sample, allows to center the fit on the RC.

The kernel $\phi$ we used corresponds to a Gaussian model of the parallax errors converted in magnitude space (we neglected here the photometric errors):

\begin{equation}
\phi = P_M (M|M_0) = P_\varpi (\varpi(M)|\varpi_0) {\partial \varpi \over \partial M}
\end{equation}

\begin{equation}
\phi(\alpha + \beta \; (\gk) - M_{\lambda}) \propto \mathcal{N} (\varpi_{\alpha, \beta, G}, \; \varpi, \; \sigma_\varpi) \; \varpi_{\alpha, \beta, m_{\lambda}}
\end{equation}

with $\varpi_{\alpha, \beta, m_{\lambda}} = 10^{(\alpha + \beta \; (\gk) - m_{\lambda} - 5)/5}$

This method allows to work directly with the parallaxes without selection in relative precision, avoiding the corresponding biases. 

In Table \ref{tab:absmagcoef} we summarise the results obtained with this method for 15 photometric bands. The initial uncertainties obtained through the maximum optimization algorithm appeared underestimated ($\sim 0.004$). Indeed we saw that by changing slightly the sample selection the results changed by more than the quoted errors. We provide in Table \ref{tab:absmagcoef} the uncertainties by Bootstrap instead. 

We have checked the degree of significance of the colour term for each relationship through a p-value test at $99\%$ confidence level. We find a marginal dependence on colour for $M_{\ks}$ (p-value of 0.004) as well as for $M_H$ (p-value of 0.002), negligible for $M_{W1}$, $M_{W2}$, $M_{W3}$ and $M_{W4}$, and an important dependence for $M_G$ and the other magnitudes. For those magnitudes for which there is no significant dependence we provide the results computed with $\beta$ fixed to zero (indicated with "-" in the table).  We also include in Table \ref{tab:absmagcoef} the results obtained for $M_{\ks}$ and $M_H$ without taking into account their marginal dependence on colour ($M_{\ks} = -1.606 \pm 0.009$, $M_H = -1.450 \pm  0.017$).

\begin{table}[ht]
\centering
\caption{Coefficients of the absolute magnitude calibrations of the RC: $M_{\lambda} = \alpha + \beta \; (\gk - 2.1)$}
\label{tab:absmagcoef}
\begin{tabular}{l|c|c|r}
  \hline
\multicolumn{1}{c|}{$\bm{M_{\lambda}}$} & {$\bm{\alpha}$} & {$\bm{\beta}$} & \multicolumn{1}{|c}{{\textbf{N}}} \\ 
  \hline  
   $M_B$  & 1.931 $\pm$ 0.009 & 2.145 $\pm$ 0.130 & 1043 \\ 
   $M_V$  & 0.855 $\pm$ 0.009 & 1.354 $\pm$ 0.126 & 1113 \\ 
   $M_{B_{\mbox{\textrm{\scriptsize T}}}}$  & 2.239 $\pm$ 0.009 & 2.397 $\pm$ 0.135 & 1190 \\ 
   $M_{V_{\mbox{\textrm{\scriptsize T}}}}$  & 0.975 $\pm$ 0.009 & 1.447 $\pm$ 0.127 & 1639 \\ 
   $M_g$  & 1.331 $\pm$ 0.056 & 1.961 $\pm$ 0.585 & 407 \\ 
   $M_r$  & 0.552 $\pm$ 0.026 & 1.194 $\pm$ 0.289 & 340 \\ 
   $M_i$  & 0.262 $\pm$ 0.032 & 0.626 $\pm$ 0.402 & 243 \\ 
   $M_G$  & 0.495 $\pm$ 0.009 & 1.121 $\pm$ 0.128 & 2482 \\ 
   $M_J$  & -0.945 $\pm$ 0.010 & 0.421 $\pm$ 0.117 & 2098 \\ 
   $M_H$  & -1.454 $\pm$ 0.018 & 0.234 $\pm$ 0.224 & 1315 \\ 
   $M_H\tablefootmark{*}$  & -1.450 $\pm$ 0.017 & - & 1315 \\ 
   $M_{\ks}$  & -1.605 $\pm$ 0.009 & 0.121 $\pm$ 0.125 & 2482 \\ 
   $M_{\ks}\tablefootmark{*}$  & -1.606 $\pm$ 0.009 & - & 2482 \\ 
   $M_{W1}$  & -1.711 $\pm$ 0.017 & - & 962 \\ 
   $M_{W2}$  & -1.585 $\pm$ 0.016 & - & 1031 \\ 
   $M_{W3}$  & -1.638 $\pm$ 0.011 & - & 2026 \\ 
   $M_{W4}$  & -1.704 $\pm$ 0.012 & - & 747 \\ 
   \hline
\end{tabular}
\tablefoot{
\tablefoottext{*}{Result without taking into account the marginal dependence on colour (p-value < 0.005)}
}
\end{table}

We checked the robustness of the mode estimate versus the selected sample. We found differences of 0.006 mag when selecting only stars with $\sigma_\varpi/\varpi<10\%$ (1085 stars).

We determined similarly the mode of the RC $M_{\ks}$ distribution according to the Padova isochrones, simulating an HR diagram with a constant Star Formation Rate (SFR), a \cite{chabrier_galactic_2001} Initial Mass Function (IMF), and a Gaussian metallicity distribution (0, 0.02). We obtained $M_{\ks} = -1.660 \pm 0.003$, in agreement with \cite{bovy_apogee_2014}. We checked on this simulation that indeed the mode is robust to changes in the SFR, the IMF and the Age-Metalliticy Ratio (AMR) hypothesis.

A summary of various absolute magnitude calibrations in the literature can be found in Table \ref{tab:mabslit}, based on Table 1 of \cite{girardi_rcstars_2016} and complemented with more recent studies. In this table we indicate, for comparison purposes, our result from Table \ref{tab:absmagcoef} assuming $\gk$ colour equal to 2.1 when the external calibrations did not consider a colour effect while we found such a dependency.
 
We found general agreement with the $M_{\ks}$ from previous works who mainly used Hipparcos data. The $M_{\ks}$ value of \cite{alves_k-band_2000} is in the TMSS system \citep{bessellBrett_jhksystem_1988}, while the others, including this work, mainly used 2MASS data. However the quality flags considered to select the data are not the same in each case. Our $M_{\ks}$ value of the mean RC K-band absolute magnitude appears to be slightly lower than in \cite{alves_k-band_2000}, \cite{grocholski_wiyn_2002} and \cite{laney_jhk_2012}, but higher than the values in \cite{vanhekshoecht_kband_2007}, \cite{groenewegen_red_2008} and \cite{francis_distancegc_2014}. It perfectly agrees with the last result of \cite{hawkins_rccalibration_2017} using also Gaia data but with a very different selection function and handling of the extinction. As in this work, \cite{groenewegen_red_2008} also found a weak dependency of $M_{\ks}$ on colour.
	
For $M_J$, \cite{laney_jhk_2012} found a slightly larger result with respect to us. However, the source of photometric data is different from ours, and we have a much larger sample. \cite{chen_absmags_2017} used a much smaller sample and their value is even higher than the one from \cite{laney_jhk_2012} but still consistent with this work. We find perfect agreement with \cite{hawkins_rccalibration_2017} who also used a sample of Gaia stars. The same authors also calibrated $M_H$, the results of which are in fair agreement with our value.

\cite{chen_absmags_2017} also calibrated the APASS-SLOAN $gri$ absolute magnitudes using seismically determined RC stars from the Strömgren survey for Asteroseismology and Galactic Archaeology (SAGA). We find that the RC is less bright in all three magnitudes although within the errors bars.

As shown in Table \ref{tab:mabslit}, for $M_{W1}$ our result agrees with both \cite{chen_absmags_2017} and \cite{hawkins_rccalibration_2017}, and it is marginally brighter than the one from \cite{yaz_wiseRCabsmags_2013}. For $M_{W2}$ we also find good agreement with \cite{chen_absmags_2017}, however the differences are larger with respect to \cite{hawkins_rccalibration_2017} as they already point out in their article. We have indeed found important variations depending on the sample selection criteria. In particular, by considering only the high photometric quality flag (i.e. qph = A) and no cut in the observed magnitude, we obtained $M_{W2} = -1.68 \pm 0.01$ with a strong correlation with colour. By removing the saturated stars \citep{cotten_comprehensiveNIR_2016}, as described in Section \ref{sect:photdata}, this dependence with colour becomes negligible, as it is for the other WISE bands, and the peak is much fainter. This may explain the too bright value found by \cite{hawkins_rccalibration_2017}.

With $M_{W3}$ all the works are consistent among them, with the exception of \cite{chen_absmags_2017} who obtained a brighter value. And for $M_{W4}$ our result is fainter than the one found for the first time by \cite{hawkins_rccalibration_2017}.

Finally, \cite{hawkins_rccalibration_2017} also provided the first calibration of $M_G$, based on a hierarchical probabilistic model. In this work we provide a different approach by directly using the mode of the distribution, and with a larger sample of data. Their $G$ absolute magnitude is somewhat brighter. As mentioned above, with our method we also find a strong dependence on colour. This may explain the difference between both estimations, together with the fact that they corrected the reddening by deriving the extinction coefficients through the nominal Gaia $G$ band. The updated extinction coefficients will be found in \cite{danielski_kG_2017}, in prep.

\begin{table*}
\caption{Comparison of the $M_{\lambda}$ of this work with other determinations in the literature}\label{tab:mabslit}
\begin{center}
\resizebox{\textwidth}{!}{\begin{tabular}{c|l|c|c|c} \hline\hline
$\bm{M_{\lambda}}$ & \multicolumn{1}{c|}{\textbf{Reference}} & \textbf{Calibration} & \multicolumn{1}{c|}{\textbf{Sample}} & \multicolumn{1}{c}{\textbf{Extinction correction}} \\\hline 
 & & & & \\[-0.5em]
 
\multirow{3}{*}{$\bm{M_G}$}
  & \cite{hawkins_rccalibration_2017} & 0.44 $\pm$ 0.01 & 972 TGAS (Gaia DR1) & $E(B-V)$ from 3D dustmap of \cite{green_3dmap_2015} \\\cline{2-5}
 & & & & \\[-0.5em]
& \multirow{2}{*}{This work} &\multirow{2}{*}{0.495 $\pm$ 0.009 \tablefootmark{*}} & \multirow{2}{*}{2482 TGAS (Gaia DR1)} & None: low extinction stars selection according \\
& & & & to \cite{capitanio_3dmap_2017} 3D map \\\hline

\multicolumn{5}{c}{} \\\hline
 & & & & \\[-0.5em]
\multirow{7}{*}{$\bm{M_J}$}
& \multirow{2}{*}{\cite{laney_jhk_2012}} &  \multirow{2}{*}{-0.984 $\pm$ 0.014} & 191 Revised Hipparcos parallaxes with SAAO $JK$ mag, & \multirow{2}{*}{None} \\
& & & data corrected for Lutz-Kelker bias & \\\cline{2-5}
 & & & & \\[-0.5em]
& \cite{chen_absmags_2017} &-1.016 $\pm$ 0.063 & $\lesssim$ 171 RC stars of the SAGA survey with 2MASS $J$ mag & SAGA $E(B-V)$ with \cite{cardelli_relationship_1989} law \\\cline{2-5}
& & & & \\[-0.5em]
& \cite{hawkins_rccalibration_2017}& -0.93 $\pm$ 0.01 & 972 TGAS (Gaia DR1) with 2MASS $J$ mag & $E_{B-V}$ from 3D dustmap of \cite{green_3dmap_2015} \\\cline{2-5}
 & & & & \\[-0.5em]
& \multirow{2}{*}{This work} &\multirow{2}{*}{-0.945 $\pm$ 0.01 \tablefootmark{*}} & \multirow{2}{*}{2098 TGAS (Gaia DR1) with 2MASS $J$ mag} & None: low extinction stars selection according \\
& & & & to \cite{capitanio_3dmap_2017} 3D map \\\hline

\multicolumn{5}{c}{} \\\hline
 & & & & \\[-0.5em]
\multirow{7}{*}{$\bm{M_H}$}
& \multirow{2}{*}{\cite{laney_jhk_2012}} &  \multirow{2}{*}{-1.490 $\pm$ 0.015} & 191 Revised Hipparcos parallaxes with SAAO $JK$ mag, & \multirow{2}{*}{None} \\
& & & data corrected for Lutz-Kelker bias & \\\cline{2-5}
 & & & & \\[-0.5em]
& \cite{chen_absmags_2017} & -1.528 $\pm$ 0.055 & $\lesssim$ 171 RC stars of the SAGA survey with 2MASS $J$ mag & SAGA $E(B-V)$ with \cite{cardelli_relationship_1989} law \\\cline{2-5}
& & & & \\[-0.5em]
& \cite{hawkins_rccalibration_2017}& -1.46 $\pm$ 0.01 & 972 TGAS (Gaia DR1) with 2MASS $J$ mag & $E_{B-V}$ from 3D dustmap of \cite{green_3dmap_2015} \\\cline{2-5}
 & & & & \\[-0.5em]
& \multirow{2}{*}{This work} &\multirow{2}{*}{-1.450 $\pm$ 0.017} & \multirow{2}{*}{1315 TGAS (Gaia DR1) with 2MASS $J$ mag} & None: low extinction stars selection according \\
& & & & to \cite{capitanio_3dmap_2017} 3D map \\\hline

\multicolumn{5}{c}{} \\\hline
 & & & & \\[-0.5em]
\multirow{16}{*}{$\bm{M_K}$}
 & \cite{alves_k-band_2000} & -1.61 $\pm$ 0.03 & 238 Hipparcos RC giants with TMSS $K$ mag & None \\\cline{2-5}
 & & & & \\[-0.5em]
& \cite{grocholski_wiyn_2002} & -1.61 $\pm$ 0.04 & 14 WYIN Open Clusters & \cite{twarog_constraints_1997} $E(B-V)$ with \cite{cardelli_relationship_1989} law \\\cline{2-5}
 & & & & \\[-0.5em]
& \cite{vanhekshoecht_kband_2007} & -1.57 $\pm$ 0.05 & 24 2MASS Open Clusters & \cite{twarog_constraints_1997} $E(B-V)$ data \\\cline{2-5}
 & & & & \\[-0.5em]
& \cite{groenewegen_red_2008} & -1.54 $\pm$ 0.04 & Revised Hipparcos parallaxes with 2MASS $K$ mag & Based on three 3D models \\\cline{2-5}
 & & & & \\[-0.5em]
& \multirow{2}{*}{\cite{laney_jhk_2012}} & \multirow{2}{*}{ -1.613 $\pm$ 0.015} & 191 Revised Hipparcos parallaxes with SAAO $K$ mag, & \multirow{2}{*}{None} \\
& & & data corrected for Lutz-Kelker bias & \\\cline{2-5}
 & & & & \\[-0.5em]
& \multirow{2}{*}{\cite{francis_distancegc_2014}} & \multirow{2}{*}{-1.53 $\pm$ 0.01} & \multirow{2}{*}{Revised Hipparcos parallaxes with 2MASS $K$ mag} & Outside 100pc: \cite{brustein_reddeningmap_1978, burstein_redmap_1982} map \\
& & & &  and \cite{bahcall_starcounts_1980} formulae \\\cline{2-5}
 & & & & \\[-0.5em]
  & \cite{chen_absmags_2017} &-1.626 $\pm$ 0.057 & $\lesssim$ 171 RC stars of the SAGA survey with 2MASS $K$ band & SAGA $E(B-V)$ with \cite{cardelli_relationship_1989} law \\\cline{2-5}
  & & & & \\[-0.5em]
 & \cite{hawkins_rccalibration_2017}& -1.61 $\pm$ 0.01 & 972 TGAS (Gaia DR1) with 2MASS $K$ mag & $E(B-V)$ from 3D dustmap of \cite{green_3dmap_2015} \\\cline{2-5}
  & & & & \\[-0.5em]
& \multirow{2}{*}{This work} &\multirow{2}{*}{-1.606 $\pm$ 0.009} & \multirow{2}{*}{2482 TGAS (Gaia DR1) with 2MASS $K$ mag} & None: low extinction stars selection according \\
& & & & to \cite{capitanio_3dmap_2017} 3D map \\\hline

\multicolumn{5}{c}{} \\\hline
 & & & & \\[-0.5em]
\multirow{3}{*}{$\bm{M_g}$}
   & \cite{chen_absmags_2017} & 1.229 $\pm$ 0.172 & $\lesssim$ 171 RC stars of the SAGA survey with APASS-SLOAN $g$ mag & SAGA $E(B-V)$ with \cite{cardelli_relationship_1989} law \\\cline{2-5}
    & & & & \\[-0.5em]
& \multirow{2}{*}{This work} &\multirow{2}{*}{1.331 $\pm$ 0.056\tablefootmark{*}} & \multirow{2}{*}{407 TGAS (Gaia DR1) with APASS-SLOAN $g$ mag} & None: low extinction stars selection according \\
& & & & to \cite{capitanio_3dmap_2017} 3D map \\\hline

\multicolumn{5}{c}{} \\\hline
 & & & & \\[-0.5em]
\multirow{3}{*}{$\bm{M_r}$}
   & \cite{chen_absmags_2017} &0.420 $\pm$ 0.110 & $\lesssim$ 171 RC stars of the SAGA survey with APASS-SLOAN $r$ mag & SAGA $E(B-V)$ with \cite{cardelli_relationship_1989} law \\\cline{2-5}
    & & & & \\[-0.5em]
& \multirow{2}{*}{This work} &\multirow{2}{*}{0.552 $\pm$ 0.026 \tablefootmark{*}} & \multirow{2}{*}{340 TGAS (Gaia DR1) with APASS-SLOAN $r$ mag} & None: low extinction stars selection according \\
& & & & to \cite{capitanio_3dmap_2017} 3D map \\\hline

\multicolumn{5}{c}{} \\\hline
 & & & & \\[-0.5em]
\multirow{3}{*}{$\bm{M_i}$}
   & \cite{chen_absmags_2017} &0.157 $\pm$ 0.094 & $\lesssim$ 171 RC stars of the SAGA survey with APASS-SLOAN $i$ mag & SAGA $E(B-V)$ with \cite{cardelli_relationship_1989} law \\\cline{2-5}
    & & & & \\[-0.5em]
& \multirow{2}{*}{This work} &\multirow{2}{*}{0.262 $\pm$ 0.032\tablefootmark{*}} & \multirow{2}{*}{243 TGAS (Gaia DR1) with APASS-SLOAN $i$ mag} & None: low extinction stars selection according \\
& & & & to \cite{capitanio_3dmap_2017} 3D map \\\hline

\multicolumn{5}{c}{} \\\hline
 & & & & \\[-0.5em]
\multirow{6}{*}{$\bm{M_{W1}}$}
   & \cite{yaz_wiseRCabsmags_2013} & -1.635 $\pm$ 0.026 & 3889 Revised Hipparcos RC parallaxes with WISE $W1$ mag & $E(B-V)$ from 2D map of \cite{schlegel_maps_1998} \\\cline{2-5}
    & & & & \\[-0.5em]
   & \cite{chen_absmags_2017} &-1.694 $\pm$ 0.061 & $\lesssim$ 171 RC stars of the SAGA survey with WISE $W1$ mag & SAGA $E(B-V)$ with \cite{cardelli_relationship_1989} law \\\cline{2-5}
    & & & & \\[-0.5em]
  & \cite{hawkins_rccalibration_2017}& -1.68 $\pm$ 0.02 & 936 TGAS (Gaia DR1) with WISE $W1$ mag & $E(B-V)$ from 3D dustmap of \cite{green_3dmap_2015} \\\cline{2-5}
 & & & & \\[-0.5em]
& \multirow{2}{*}{This work} &\multirow{2}{*}{-1.711 $\pm$ 0.017} & \multirow{2}{*}{962 TGAS (Gaia DR1) with WISE $W1$ mag} & None: low extinction stars selection according \\
& & & & to \cite{capitanio_3dmap_2017} 3D map \\\hline

\multicolumn{5}{c}{} \\\hline
 & & & & \\[-0.5em]
\multirow{4}{*}{$\bm{M_{W2}}$}
   & \cite{chen_absmags_2017} &-1.595 $\pm$ 0.064 & $\lesssim$ 171 RC stars of the SAGA survey with WISE $W2$ mag & SAGA $E(B-V)$ with \cite{cardelli_relationship_1989} law \\\cline{2-5}
    & & & & \\[-0.5em]
  & \cite{hawkins_rccalibration_2017} & -1.69 $\pm$ 0.02 & 934 TGAS (Gaia DR1) with WISE $W2$ mag & $E(B-V)$ from 3D dustmap of \cite{green_3dmap_2015} \\\cline{2-5}
 & & & & \\[-0.5em]
& \multirow{2}{*}{This work} &\multirow{2}{*}{-1.585 $\pm$ 0.016} & \multirow{2}{*}{1031 TGAS (Gaia DR1) with WISE $W2$ mag} & None: low extinction stars selection according \\
& & & & to \cite{capitanio_3dmap_2017} 3D map \\\hline

\multicolumn{5}{c}{} \\\hline
 & & & & \\[-0.5em]
\multirow{6}{*}{$\bm{M_{W3}}$}
   & \cite{yaz_wiseRCabsmags_2013} & -1.606 $\pm$ 0.024 & 3889 Revised Hipparcos RC parallaxes with WISE $W3$ mag & $E(B-V)$ from 2D map of \cite{schlegel_maps_1998} \\\cline{2-5}
    & & & & \\[-0.5em]
   & \cite{chen_absmags_2017} &-1.752 $\pm$ 0.068 & $\lesssim$ 171 RC stars of the SAGA survey with WISE $W3$ mag & SAGA $E(B-V)$ with \cite{cardelli_relationship_1989} law \\\cline{2-5}
    & & & & \\[-0.5em]
  & \cite{hawkins_rccalibration_2017}& -1.67 $\pm$ 0.02 & 936 TGAS (Gaia DR1) with WISE $W3$ mag & $E(B-V)$ from 3D dustmap of \cite{green_3dmap_2015} \\\cline{2-5}
 & & & & \\[-0.5em]
& \multirow{2}{*}{This work} &\multirow{2}{*}{-1.638 $\pm$ 0.011} & \multirow{2}{*}{2026 TGAS (Gaia DR1) with WISE $W3$ mag} & None: low extinction stars selection according \\
& & & & to \cite{capitanio_3dmap_2017} 3D map \\\hline

\multicolumn{5}{c}{} \\\hline
 & & & & \\[-0.5em]
\multirow{3}{*}{$\bm{M_{W4}}$}
  & \cite{hawkins_rccalibration_2017}& -1.76 $\pm$ 0.01 & 910 TGAS (Gaia DR1) with WISE $W4$ mag & $E(B-V)$ from 3D dustmap of \cite{green_3dmap_2015} \\\cline{2-5}
 & & & & \\[-0.5em]
& \multirow{2}{*}{This work} &\multirow{2}{*}{-1.704 $\pm$ 0.012} & \multirow{2}{*}{747 TGAS (Gaia DR1) with WISE $W4$ mag} & None: low extinction stars selection according \\
& & & & to \cite{capitanio_3dmap_2017} 3D map \\\hline
\end{tabular}}
\tablefoot{
\tablefoottext{*}{Result from Table \ref{tab:absmagcoef} assuming $\gk$ colour equal to 2.1}
}
\end{center}
\end{table*}

Besides the parallax accuracy and the various sources of photometric information, one of the main differences between these estimates is the handling of the interstellar extinction: \cite{alves_k-band_2000}, \cite{stanek_distance_1998}, \cite{girardi_fine_1998} and \cite{laney_jhk_2012} assumed no reddening, while the other authors corrected their magnitudes using and combining different interstellar laws and/or maps. It is clear that our sample selection based on low extinction stars introduces as well a bias in all our calibrations, although minimally. Indeed, the reddening cut at $E(B-V)_{\rm max} < $ 0.01 corresponds to a maximum over-estimation of the absolute magnitude of about 0.02 mag in the $G$ band, while about 0.003 mag in the $K$ band \citep[see][in prep.]{danielski_kG_2017}.

The discrepancies among the other estimates may also be justified by the different methods used: most of the authors considered a Gaussian fit, while here we used the mode of the distribution.


\section{The TGAS RC HRD}  \label{sect:applications}

\begin{figure}
   \centering
   \includegraphics[width=0.50\textwidth]{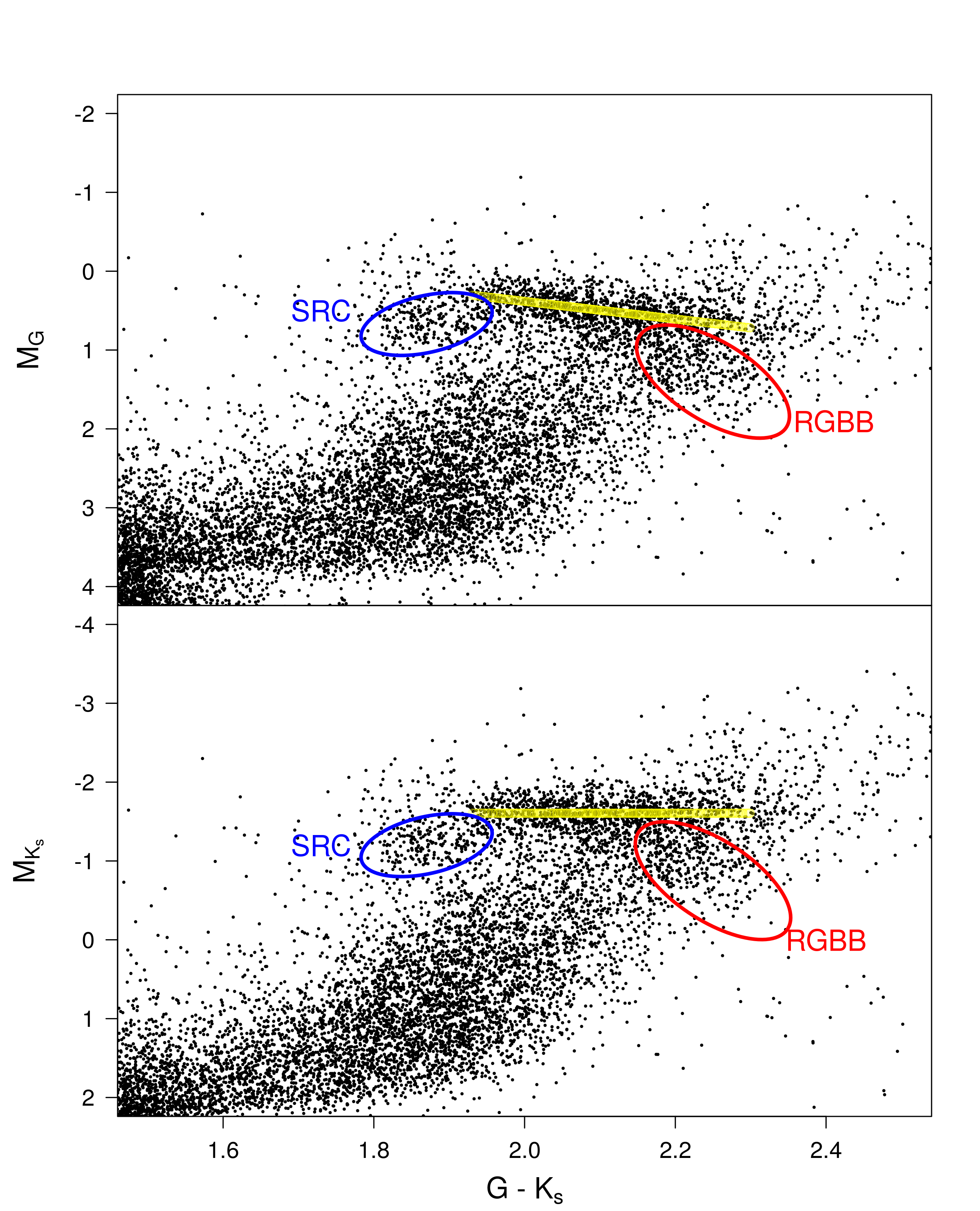}
      \caption{TGAS HR diagram of the RC region for the $M_G$ (top) and the $M_{\ks}$ (bottom) absolute magnitudes, using stars with $E(B-V)_{\rm max} < 0.015$, $10\%$ parallax precision, $\sigma_G < 0.01$ and 2MASS $JK$ high photometric quality. The location of the Secondary Red Clump (SRC) and the Red Giant Branch Bump (RGBB) features are easily observed on the diagram. We have highlighted them in blue and red, respectively. The yellow line shows the absolute magnitude calibration obtained in this work}
         \label{hrtgas}
   \end{figure}

Figure \ref{hrtgas} shows the TGAS HR diagram for red giant stars for absolute magnitudes in the $G$ and $K$ photometric bands. We used stars listed in Table \ref{tab:hrtgas} (see Appendix \ref{annex:hrtgas}), with low extinction ($E(B-V)_{\rm max} < 0.015$), $10\%$ parallax precision, $\sigma_G < 0.01$ and 2MASS $JK$ high photometric quality. In  both cases the RC is easily detected. However other features may also be observed. Indeed, on the bluest part of the RC we can see a small overdensity belonging to the Secondary Red Clump \citep{girardi_fine_1998, girardi_clump_1999}, a group of still metal-rich but younger (i.e. slightly more massive) stars that extend the RC to fainter magnitudes (up to 0.4 mag fainter). On the red side of the clump and below it, we find the Red Giant Branch Bump (RGB bump or RGBB), another overdensity of slightly more massive stars than the RC which causes a peak (\textit{bump}) in the luminosity function (see \cite{christensen_onthered_2015} for a review on this CMD feature).

On the same diagrams we also overplotted the absolute magnitude calibrations obtained in previous Sect. \ref{sect:absmag} (Table \ref{tab:absmagcoef}).

\begin{figure}
   \centering
   \includegraphics[width=0.50\textwidth]{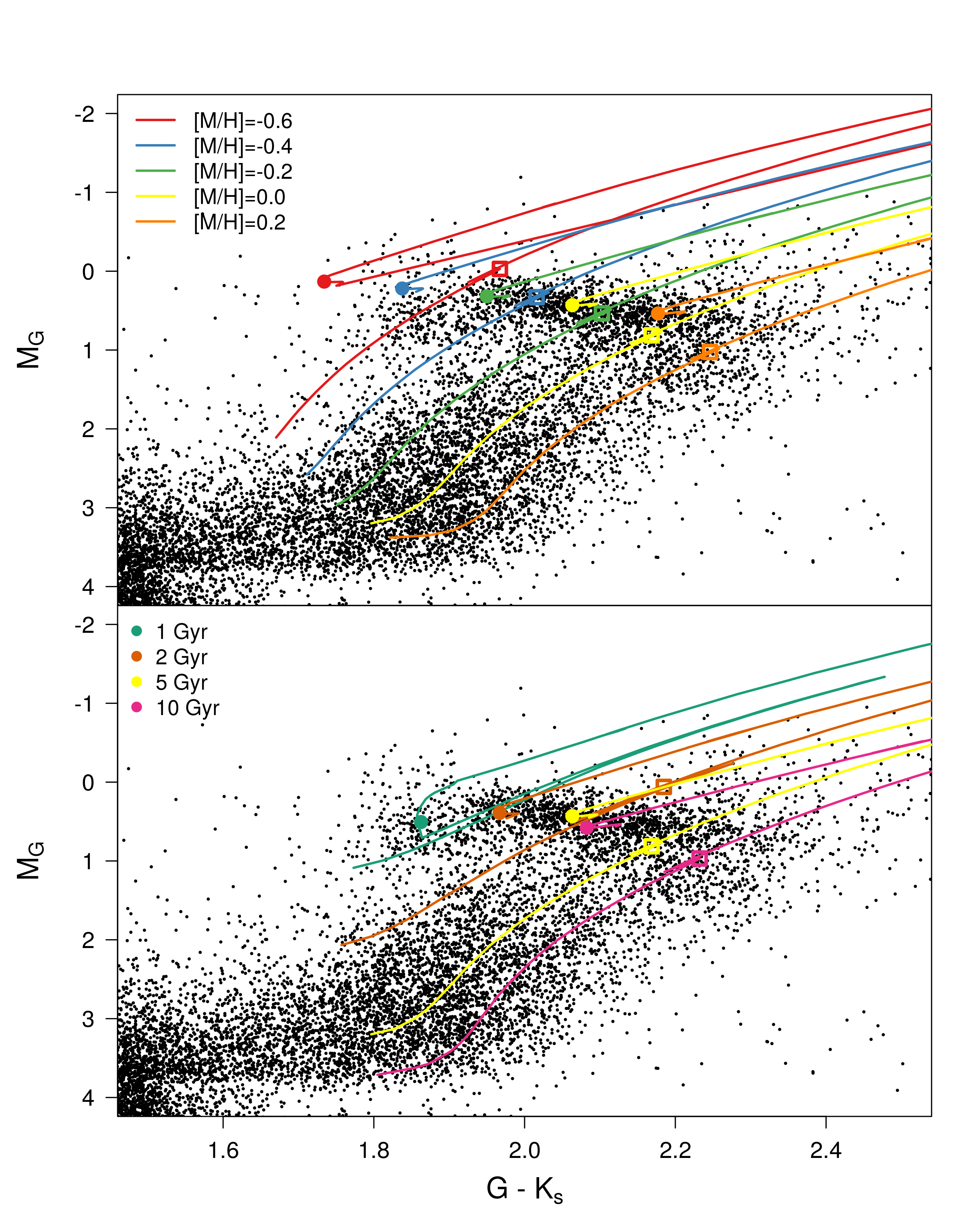}
      \caption{Padova isochrones overlaid on the Gaia DR1 HRD. Top: isochrones for an age of 5 Gyr and different metallicities. Bottom: isochrones for a solar metallicity and different ages. Circles correspond to the Red Clump location, squares to the RGB bump}
         \label{isocs}
   \end{figure}

In Fig. \ref{isocs} we show again the RC HR diagram but now overplotting in different colours the Padova isochrones \cite[][Parsec 2.7]{bressan_isopadova_2012} at different metallicities (top panel) and at different ages (bottom panel). We use the original $\teff$ from the isochrones and applied our $\gk$ vs $\tnorm$ calibration (Table \ref{tab:teffccoeffs}) to derive the colour $\gk$.

We may see that while the position of the RC seems to nicely fit the isochrones, the RGB bump is slightly too bright in the isochrones. Following the process in Sect. \ref{sect:absmag}, we found a difference of -0.07 mag between the Padova RC and the TGAS RC, and about 0.2 mag for the RGB bump.

\section{Conclusions} \label{sect:conclusions}

Using the First Gaia Data Release parallaxes and photometry, the new 3D interstellar extinction map of \cite{capitanio_3dmap_2017}, the 2MASS catalogue and the last APOGEE release (DR13), a complete photometric calibration including colours (spread from visual to infrared wavelengths), absolute magnitudes, spectroscopic metallicities, and homogeneous effective temperatures, is provided in this work for solar neighbourhood RC stars. We have made use of high photometric quality data from the Gaia, Johnson, 2MASS, Hipparcos, Tycho-2 and APASS-SLOAN and WISE photometric systems. 

A robust MCMC method accounting for all variable uncertainties was developed to derive 20 accurate metallicity-dependent colour-colour relations and the $\tnorm$ vs $\gk$ and $\gk$ vs $\tnorm$ fits (with $\tnorm$ the normalized effective temperature, $\tnorm = \teff / 5040$). We checked that the effective temperature calibration is compatible with those from \cite{ramirez_effective_2005}, \cite{gonzalez_hernandez_new_2009} and \cite{huang_empirical_2015} within the metallicity and colour ranges of applicability.

We also derived the absolute magnitudes for the TGAS RC on 15 photometric bands (including $M_G$ and $M_{\ks}$) through a kernel based magnitude distribution method, and using the largest high quality dataset used so far for an absolute magnitude calibration of the RC. We obtained a small dependence on colour for $M_{\ks}$ and $M_H$, not-significant for $M_{W1}$, $M_{W2}$, $M_{W3}$ and $M_{W4}$, but important for $M_G$ and the other magnitudes. 

Note that all these photometric relationships will be improved in later Gaia releases as well as extended to other photometric bands, when larger Red Clump samples will be available.

We presented an un-reddened TGAS HR diagram for the RC region, in which we can already easily identify other features of red giant stars, such as the Secondary Red Clump and the Red Giant Branch Bump. By using our calibrations we could compare the Padova isochrones with the TGAS HR diagram and found good agreement with the RC location on the diagram, although the RGB bump appears too bright in the isochrones. 

The photometric calibrations presented here are being used to derive $k_G$, the interstellar extinction coefficient in the $G$-band \citep[][in prep.]{danielski_kG_2017}, and to provide photometric interstellar extinctions of large surveys like APOGEE to be included in the next release of the new 3D extinction map of \cite{capitanio_3dmap_2017}.

In summary, this work used the high quality of the Gaia DR1 data to calibrate the Gaia Red Clump. In turn, these calibrations can be used as the second rung of the cosmic distance ladder. Indeed, together with asteroseismic constraints, we can now derive the distance modulus of a large sample of RC stars. By choosing RC stars distant enough so that their estimated distance uncertainty is better than the Gaia parallax precision, these stars may be used to check the zero point of the Gaia parallaxes and their precision \citep{arenou_cu9validationDR1_2017}. This is already being applied within the Gaia Data Release 2 verification process.

\vspace{0.5cm}


\begin{acknowledgements}
L.R.-D. acknowledges financial support from the \textit{Centre National d’Etudes Spatiales} (CNES) fellowship program. This work has made use of data from the European Space Agency (ESA) mission {\it Gaia} (\url{https://www.cosmos.esa.int/gaia}), processed by the {\it Gaia} Data Processing and Analysis Consortium (DPAC, \url{https://www.cosmos.esa.int/web/gaia/dpac/consortium}). Funding for the DPAC has been provided by national institutions, in particular the institutions participating in the {\it Gaia} Multilateral Agreement. This publication makes use of data products from the Two Micron All Sky Survey, which is a joint project of the University of Massachusetts and the Infrared Processing and Analysis Center/California Institute of Technology, funded by the National Aeronautics and Space Administration and the National Science Foundation. L.R.-D. acknowledges support from \textit{Agence Nationale de la Recherche} through the STILISM project (ANR-12-BS05-0016-02). This research has also made use of VizieR databases operated at the \textit{Centre de Données astronomiques de Strasbourg} (CDS) in France.
\end{acknowledgements}


\bibliographystyle{aa}
\bibliography{References}

\begin{appendix}

\section{Low extinction TGAS HR Catalogue at CDS} \label{annex:hrtgas}

Table \ref{tab:hrtgas} contains a few rows of the low extinction TGAS HRD compilation used in this work. The full table is available in VizieR.

The catalogue includes 142996 stars with:

\begin{itemize}
\item Gaia DR1 and 2MASS identifiers
\item Gaia DR1 parallaxes with precision better than 10$\%$
\item Gaia DR1 $G$ magnitude with uncertainties lower than 0.01 mag
\item 2MASS $J$ and $\ks$ photometric bands with high quality (i.e. flag q2M = "A.A") and uncertainties lower than 0.03 mag
\item Reddening $E(B-V)_{\rm max} < 0.015$ according to the \cite{capitanio_3dmap_2017} 3D interstellar extinction map, and the \cite{schlegel_maps_1998} 2D map for stars for which the distance go beyond the 3D map borders
\end{itemize}

\begin{table*}
\centering
\caption{First rows of the low extinction and high photometric and astrometric quality TGAS HRD catalogue} 
\label{tab:hrtgas}
  \resizebox{\textwidth}{!}{\begin{tabular}{c|c|c|c|c|c|c|c|c|c|c} \hline\hline  
\textbf{GDR1 id} & \textbf{2MASS id} & $\bm{\varpi}$ & $\bm{\sigma_{\varpi}}$ & $\bm{G}$ & $\bm{\sigma_{G}}$  & $\bm{J_s}$ & $\bm{\sigma_{J_s}}$ & $\bm{K_s}$ & $\bm{\sigma_{K_s}}$ &  $\bm{E(B-V)_{\rm max}}$ \\\hline
7627862074752 & 03000819+0014074  & 6.353 & 0.308 & 7.991 & 0.001 & 6.606 & 0.023 & 6.019 & 0.020 & 0.011 \\ 
  16527034310784 & 03003397+0021355  & 8.663 & 0.256 & 9.972 & 0.001 & 8.993 & 0.018 & 8.651 & 0.025 & 0.004 \\ 
  26834955821312 & 03000244+0021039  & 6.202 & 0.247 & 9.971 & 0.001 & 9.189 & 0.023 & 8.860 & 0.025 & 0.012 \\ 
  44358422235136 & 03020031+0029521  & 9.958 & 0.548 & 9.317 & 0.004 & 8.332 & 0.023 & 7.990 & 0.024 & 0.003 \\ 
  115723598973952 & 03002534+0048455  & 10.550 & 0.232 & 10.788 & 0.000 & 9.502 & 0.022 & 8.921 & 0.020 & 0.003 \\ 
  122732985598464 & 03004702+0059362  & 6.582 & 0.303 & 8.774 & 0.001 & 8.071 & 0.026 & 7.833 & 0.020 & 0.010 \\ 
  308619170261760 & 02572363+0058185  & 7.446 & 0.279 & 10.465 & 0.001 & 9.437 & 0.026 & 8.969 & 0.023 & 0.006 \\ 
  310337157179392 & 02572548+0059538  & 7.381 & 0.247 & 10.592 & 0.001 & 9.513 & 0.023 & 9.102 & 0.021 & 0.007 \\ 
  320713798164992 & 02585888+0104389  & 7.481 & 0.284 & 9.993 & 0.002 & 9.094 & 0.026 & 8.728 & 0.020 & 0.007 \\ 
  349369819955456 & 02580800+0122163  & 6.770 & 0.261 & 9.469 & 0.001 & 8.573 & 0.026 & 8.255 & 0.027 & 0.009 \\ 
   ... & $...$  & $...$ & $...$ & $...$ & $...$ & $...$ & $...$ & $...$ & $...$ & $...$ \\\hline
\end{tabular}}
\end{table*}

\end{appendix}
 
\end{document}